\documentclass[aps,prd,twocolumn,groupedaddress,amssymb,eqsecnum,showpacs,epsfig]{revtex4}
\usepackage{graphicx}
\usepackage{bm}
\usepackage{dcolumn}
\usepackage{amsmath}
\def \lleq {\lower0.9ex\hbox{ $\buildrel < \over \sim$} ~}
\def \ggeq {\lower0.9ex\hbox{ $\buildrel > \over \sim$} ~}

\def \omx  {\Omega_{0X}}

\def \omb   {\Omega_b}
\def \omk   {\Omega_k}
\def \omr   {\Omega_r}
\def \omm  {\Omega_{0 {\rm m}}}
\def \oml   {\Omega_{\Lambda}}

\def \beq  {\begin{equation}}
\def \eeq  {\end{equation}}
\def \ber  {\begin{eqnarray}}
\def \eer  {\end{eqnarray}}

\def\apj{{Astroph.\@ J.\ }}

\def\aj{{Astron.\@ J.\ }}

\begin{document}
\newcommand{\newc}{\newcommand}

\newc{\be}{\begin{equation}}
\newc{\ee}{\end{equation}}
\newc{\ba}{\begin{eqnarray}}
\newc{\ea}{\end{eqnarray}}
\newc{\bea}{\begin{eqnarray*}}
\newc{\eea}{\end{eqnarray*}}
\newc{\D}{\partial}
\newc{\ie}{{\it i.e.} }
\newc{\eg}{{\it e.g.} }
\newc{\etc}{{\it etc.} }
\newc{\etal}{{\it et al.}}
\newcommand{\nn}{\nonumber}
\newc{\ra}{\rightarrow}
\newc{\lra}{\leftrightarrow}
\newc{\lsim}{\buildrel{<}\over{\sim}}
\newc{\gsim}{\buildrel{>}\over{\sim}}
\title{Crossing the Phantom Divide: Theoretical Implications and
Observational Status}
\author{S. Nesseris$^a$ and L. Perivolaropoulos$^b$ }
\affiliation{Department of Physics, University of Ioannina, Greece
\\ $^a$ e-mail: me01629@cc.uoi.gr, $^b$ e-mail:
leandros@cc.uoi.gr}
\date {\today}

\begin{abstract}
If the dark energy equation of state parameter $w(z)$ crosses the
phantom divide line $w=-1$ (or equivalently if the expression
$\frac{d(H^2(z))}{dz}-3\Omega_{0m} H_0^2 (1+z)^2$ changes sign) at
recent redshifts, then there are two possible cosmological
implications: Either the dark energy consists of multiple components
with at least one non-canonical phantom component or general
relativity needs to be extended to a more general theory on
cosmological scales. The former possibility requires the existence
of a phantom component which has been shown to suffer from serious
theoretical problems and instabilities. Therefore, the latter
possibility is the simplest realistic theoretical framework in which
such a crossing can be realized.  After providing a pedagogical
description of various dark energy observational probes, we use a
set of such probes (including the Gold SnIa sample, the first year
SNLS dataset, the 3-year WMAP CMB shift parameter, the SDSS baryon
acoustic oscillations peak (BAO), the X-ray gas mass fraction in
clusters and the linear growth rate of perturbations at $z=0.15$ as
obtained from the 2dF galaxy redshift survey) to investigate the
priors required for cosmological observations to favor crossing of
the phantom divide. We find that a low $\omm$ prior ($0.2<\omm
<0.25$) leads, for most observational probes (except of the SNLS
data), to an increased probability (mild trend) for phantom divide
crossing. An interesting degeneracy of the ISW effect in the CMB
perturbation spectrum is also pointed out.
\end{abstract}
\pacs{98.80.Es,98.65.Dx,98.62.Sb}
\maketitle

\section{Introduction}

The assumption of large scale homogeneity and isotropy of the
universe combined with the assumption that general relativity is the
correct theory on cosmological scales leads to the Friedman equation
which in a flat universe takes the form \be H^2(a)=\left(\frac{{\dot
a}}{a}\right)^2=\frac{8\pi G}{3}\rho(a) \label{fe1}\ee where $a(t)$
is the scale factor of the universe and $\rho$ its average energy
density. Both sides of this equation can be observationally probed
directly: The left side using mainly geometrical methods (measuring
the luminosity and angular diameter distances
$d_L(z)$\cite{snobs,Riess:2004nr,Astier:2005qq} and
$d_A(z)$\cite{Eisenstein:2005su,Allen:2002sr,Allen:2004cd,Spergel:2006hy}
with standard candles and standard rulers) showing an accelerating
expansion at recent redshifts
 and the matter - radiation density part
of the right side using dynamical and other methods (cosmic
microwave background \cite{Spergel:2006hy}, large scale structure
observations\cite{Verde:2001sf}, lensing\cite{Hoekstra:2002xs}
etc). These observations have indicated\cite{Sahni:1999gb} that
the two sides of the Friedman equation (\ref{fe1}) can not be
equal if $\rho(a)=\rho_m(a)\sim a^{-3}$ even if a non-zero
curvature is assumed. There are two possible resolutions to this
puzzle: Either modify the right side of the Friedman equation
(\ref{fe1}) introducing a new form of {\it `dark' energy} ideal
fluid component ($\rho(a)=\rho_m(a)+\rho_X (a)$) with suitable
evolution in order to restore the equality or modify both sides by
changing the way energy density affects geometry thus modifying
the Einstein equations.

In the first class of approaches the required gravitational
properties of dark energy (see
\cite{Copeland:2006wr,Padmanabhan:2006ag,Perivolaropoulos:2006ce,Straumann:2006tv,Uzan:2006mf,Sahni:2006pa}
for recent reviews) needed to induce the accelerating expansion are
well described by its equation of state
$w(z)=\frac{p_X(z)}{\rho_X(z)}$ which enters in the second Friedman
equation as \be \frac{{\ddot a}}{a}=-\frac{4\pi G}{3}(\rho_m +
\rho_X (1+3w)) \label{fe2} \ee implying that a negative pressure
($w<-1/3$) is necessary in order to induce accelerating expansion.
The simplest viable example of dark energy is the cosmological
constant\cite{Sahni:1999gb,Peebles:2002gy,Carroll:2000fy} ($w=-1$).
This example however even though consistent with present data lacks
physical motivation. Questions like `What is the origin of the
cosmological constant?' or `Why is the cosmological constant
$10^{120}$ times smaller than its natural scale so that it starts
dominating at recent cosmological times (coincidence problem)?'
remain unanswered\cite{Weinberg:2000yb}. Attempts to replace the
cosmological constant by a dynamical scalar field
(quintessence\cite{Caldwell:1997ii,Zlatev:1998tr,Ratra:1987rm}) have
created a new problem regarding the initial conditions of
quintessence which even though can be resolved in particular cases
(tracker quintessence), can not answer the above questions in a
satisfactory way.

The parameter $w(z)$ determines not only the gravitational
properties of dark energy but also its evolution. This evolution is
easily obtained from the energy momentum conservation  \be d(\rho_X
a^3)=-p_X d(a^3) \label{econs1} \ee which leads to \be \rho_X =
\rho_{0X} e^{-3\int_1^a \frac{da'}{a'}(1+w(a'))}=\rho_{0X}
e^{3\int_0^z \frac{dz'}{1+z'}(1+w(z'))} \label{dev1} \ee Therefore
the determination of $w(z)$ is equivalent to that of $\rho_X(z)$
which in turn is equivalent to the observed $H(z)$ from the Friedman
equation (\ref{fe1}) expressed as \be H(z)=H_0 [\omm (1+z)^3 + \omx
e^{3\int_0^z \frac{dz'}{1+z'}(1+w(z'))} ] \label{fe1a} \ee Thus,
knowledge of $\omm$ and $H(z)$ suffices to determine $w(z)$ which is
obtained from equation (\ref{fe1a}) as \cite{Huterer:2000mj} \be
w(z)=\frac{\frac{2}{3}(1+z)\frac{d ln H}{dz} -1}{1-\frac{H_0^2}{H^2}
\omm (1+z)^3} \label{wzh1} \ee

In the second class of approaches the Einstein equations get
modified and the new equations combined with the assumption of
homogeneity and isotropy lead to a generalized Friedman equation
of the form \be f(H^2)=g(\rho_m) \label{modfe1} \ee where $f$ and
$g$ are appropriate functions determined by the modified gravity
theory\cite{Boisseau:2000pr,Nojiri:2006je,Sahni:2002dx,Chimento:2006ac,Carroll:2004de,Deffayet:2000uy}.
In this class of models, the parameter $w(z)$ can also be defined
from equation (\ref{wzh1}) but it can not be interpreted as
$\frac{p_X}{\rho_X}$ of a perfect fluid.

The simplest (but quite general) examples of modified gravity
theories are scalar tensor
theories\cite{Boisseau:2000pr,Esposito-Farese:2000ij,Torres:2002pe,Gannouji:2006jm,Perivolaropoulos:2005yv}
where the Newton's constant $G$ is promoted to a function of a
field $\Phi$: $8\pi G \rightarrow \frac{1}{F(\Phi)}$ whose
dynamics at the Lagrangian level is determined by a potential
$U(\Phi)$. Assuming homogeneity and isotropy, the modified
Friedman equation in these theories take the form \be H^2 =
\frac{1}{3F}(\rho_m +\frac{1}{2}{\dot \Phi}^2 + U - 3H{\dot F})
\label{fest1} \ee
%\ba H^2 &=& \frac{1}{3F}(\rho_m +\frac{1}{2}{\dot \Phi}^2 + U -
%3H{\dot F}) \label{fest1} \\ {\dot H}=-\frac{1}{2F}(\rho_m +{\dot
%\Phi}^2 +{\ddot F}-H{\dot F} \label{fest2} \ea
This equation reduces to a regular minimally coupled scalar field
dark energy (quintessence) in the general relativity limit of a
constant $F=\frac{1}{8\pi G}$. The positive nature of the kinetic
term ${\dot \Phi}^2$ however implies that certain types of
behaviors of $H$ may not be reproducible without invoking a
time-dependent $F$. These types of behavior of $H(z)$ which
include a $w(z)$ crossing the Phantom Divide Line (PDL) $w=-1$ are
potential signatures of extended gravity theories and will be
discussed in the next section.

To identify this type of signatures, a detailed form of the
observed $H(z)$ is required which may be obtained by a combination
of multiple dark energy probes. Such probes may be divided in two
classes\cite{Bertschinger:2006aw} according to the methods used to
obtain $H(z)$.
\begin{itemize}
\item {\it Geometric methods} probe the large scale geometry of
space-time directly through the redshift dependence of cosmological
distances  ($d_L(z)$ or $d_A(z)$). They thus determine $H(z)$
independent of the validity of Einstein equations. \item {\it
Dynamical methods} determine $H(z)$ by measuring the evolution of
energy density (background or perturbations) and using a gravity
theory to relate them with geometry ie with $H(z)$. These methods
rely on knowledge of the dynamical equations that connect geometry
with energy and may therefore be used in combination with geometric
methods to test these dynamical equations. \end{itemize} Examples of
geometric probes include \begin{enumerate}\item The measured
supernova distance redshift relation $d_L
(z)$\cite{snobs,Riess:2004nr,Astier:2005qq} which for a flat
universe, is connected to $H(z)$ as \be d_L(z)=(1+z)\int_0^z
\frac{dz'}{H(z')} \label{dlh1} \ee
\item The measured\cite{Spergel:2006hy,Wang:2006ts} angular
diameter distance $d_A(z_{rec})$ to the sound horizon $r_s(z_{rec})$
at recombination \be
d_A(z_{rec})=\frac{1}{1+z_{rec}}\int_0^{z_{rec}}\frac{dz'}{H(z')}
\ee
\item The scale of the sound horizon measured at more recent
redshifts ($z_{BAO}$) through large scale structure redshift survey
correlation functions\cite{Eisenstein:2005su} \be D_V
(z)=\left[\left(\int_0^{z_{BAO}} \frac{dz}{H(z)}\right)^2
\frac{z_{BAO}}{H(z_{BAO})}\right]^{1/3} \label{dv1} \ee \item The
cluster gas mass fraction defined as \be
f_{gas}=\frac{M_{gas}}{M_{tot}} \label{fgdef} \ee assumed to be
constant for all clusters and proportional to $\frac{\omb}{\omm}$
probes the angular diameter
distance\cite{Allen:2002sr,Allen:2004cd,Sasaki:1996ss} to each
cluster as discussed in detail in section III.\end{enumerate} An
example of a dynamical probe of geometry is the measured linear
growth factor of the matter density perturbations $D(a)$ defined as
\be D(a)\equiv \frac{\frac{\delta \rho}{\rho}(a)}{\frac{\delta
\rho}{\rho}(a=1)} \label{dadef} \ee The measurements of $D(a)$ can
be made by several methods including the redshift distortion factor
in redshift surveys\cite{Hawkins:2002sg}, weak
lensing\cite{Fischer:1999zx}, number counts of galaxy
clusters\cite{Borgani:2001bb}, Integrated Sachs-Wolfe (ISW)
effect\cite{Pogosian:2005ez} and large scale structure power
spectrum\cite{Tegmark:2003ud,Tegmark:2003uf}. The theoretical
prediction of the evolution of $D(a)$ on sub-Hubble scales is
obtained from the Euler and matter stress energy conservation
equations as\cite{Uzan:2006mf,Stabenau:2006td} \ba
D''(k,a)&+&\left(\frac{3}{a}+\frac{H'(a)}{H(a)}\right)D'(k,a)-\nn
\\ &-&\frac{3}{2}\frac{\Omega_{0m}}{a^5 H(a)^2}f(k,a)D(k,a)=0
\label{greq}\ea with initial conditions $D(a)\simeq a$ for $a\simeq
0$. In equation (\ref{greq}) we have ignored anisotropic stresses
and dark energy perturbations\cite{Uzan:2006mf} which are expected
to have a small effect on sub-Hubble scales. The last term of
equation (\ref{greq}) emerges by connecting the metric perturbation
with the matter density perturbations. It therefore depends on the
particular form of the dynamical equations of the gravity theory
considered. This dependence is expressed through the function
$f(k,a)$ which in the case of general relativity is unity
($f(k,a)=1$) while in extended gravity theories it can take values
different from one which can even depend on the scale $k$
\cite{Stabenau:2006td}.

For example for scalar-tensor theories we have \be f(k,a) =
\frac{G_{eff}(a)}{G_{eff}(a=1)}\left(1+\frac{1}{1+\frac{k}{ma}}\right)\simeq\frac{F_0}{F(a)}
\ee where $G_{eff}(a)$ is the effective Newton's constant when the
scale factor is $a$ and $a=1$ corresponds to the present value of
the scale factor while $m$ is the mass of the scalar field $\Phi$
inducing a Yukawa cutoff to the gravitational field. In what follows
(Fig. 1) we will use a simple ansatz for the effective Newton's
constant \be \frac{G_{eff}(a)}{G_{eff}(a=1)}=1+\xi(1-a)^2
\label{geff} \ee and neglect the mass $m$ of the scalar field.
\begin{figure}
\hspace{0pt}\rotatebox{0}{\resizebox{.5\textwidth}{!}{\includegraphics{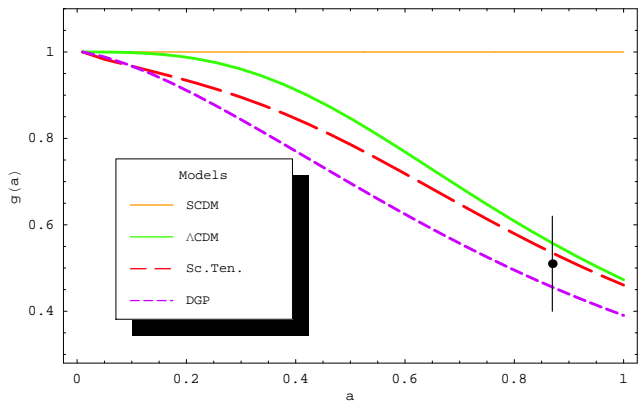}}}
\vspace{-20pt}{\caption{The growth rate in the (flat and
$\Omega_{0m}=0.26$) $\Lambda$CDM (continuous line), Scalar-Tensor
(long-dashed line) and DGP (short-dashed line) cases and a datapoint
from the 2dFGRS $g(a=0.15)=0.51\pm 0.11$. The $\Lambda$CDM model is
consistent with the current growth rate observations.}} \label{fig1}
\end{figure}
Alternatively, for the DGP model\cite{Dvali:2000rv,Deffayet:2000uy}
we have\cite{Koyama:2005kd} \be f(k,a)=(1+\frac{1}{3\beta})\ee with
\be \beta = 1-\frac{H(a)}{H_0
\sqrt{\Omega_{r_c}}}\left(1+\frac{a}{3}\frac{H'(a)}{H(a)}\right) \ee
where $r_c$  is the crossover scale beyond which the gravitational
force follows the 5-dimensional $1/r^3$ behavior \be \label{omegarc}
\Omega_{r_c} \equiv 1/4r_c^2H_0^2 \ee and $H(a)$ for the DGP model
is defined through \cite{Koyama:2005kd} \be
H_{DGP}(a)^2-\frac{H_{DGP}(a)}{r_c} =\frac{8 \pi G}{3} \rho
\label{hdgp0} \ee Solving for $H_{DGP}(a)$, assuming flatness and in
the case when we only have matter on the brane we get \be
\label{hdgp} \frac{H_{DGP}(a)}{H_0} =
\sqrt{\Omega_{r_c}}+\sqrt{\Omega_{r_c}+\omm a^{-3}}\ee where now \be
\label{omegarc1} \Omega_{r_c} = \frac {1}{4}(1-\omm)^2 \ee

The detection of an $f(k,a)\neq 1$ from equation (\ref{greq})
would therefore be a `smoking gun' signature of extended gravity
theories. Such a detection could be made for example by using the
form of $H(z)$ obtained from geometric tests in equation
(\ref{greq}), solving for $D(a)$ in the context of general
relativity ($f(k,a)=1$) and comparing with the observed $D(a)$ at
various redshifts. If a statistically significant difference is
found between the observed $D(a)$ and one predicted in the context
of general relativity then this could be interpreted as evidence
for extensions of general relativity.

There is currently an observational estimate of the growth rate
defined as \be g(a)\equiv \frac{a D'(a)}{D(a)} \label{grrat1} \ee at
a redshift $z=\frac{1}{a}-1=0.15$ from the
2dFGRS\cite{Verde:2001sf,Hawkins:2002sg,Wang:2004py}. It is \be
g(z=0.15)=0.51\pm 0.11 \label{gobs1} \ee and its derivation used the
redshift distortion factor (see section III E). There will soon be
better estimates coming from the SDSS\cite{Tegmark:2003ud}. In Fig.
1 we show the growth rate for the following flat models: a best fit
($\omm =0.26$) $\Lambda$CDM model, a DGP ($\omm =0.22$) model and a
Scalar-Tensor model ($\xi=-0.2$ in equation (\ref{geff}) and
$\omm=0.26$). As demonstrated in Fig. 1 however, the large errorbars
in the currently available datapoint of equation (\ref{gobs1}) can
not distinguish the best fit $\Lambda$CDM model (flat $\oml=0.74$)
from other competing models.
We therefore conclude that several different models including
modified gravity models as well as $\Lambda$CDM are in agreement
with current data. %%

Given the current uncertainties in the growth rate observations,
it is important to uncover potential signatures of extended
gravity theories in geometric observational methods. As discussed
in the next section, a minimally coupled scalar field dark energy
can not reproduce a $w(z)$ crossing the PDL ($w=-1$) for any
scalar field potential (see however Ref \cite{Tsujikawa:2005cd}
for an interesting case). The simplest realistic theoretical model
which can reproduce such crossing is a scalar-tensor extension of
general relativity. Therefore, if such crossing is confirmed by
geometric observations then this could be interpreted as an
indication for extended gravity theories.

The structure of this paper is the following: In the next section
we discuss the theoretical consequences emerging if the crossing
of the PDL is confirmed by future observations. In section III we
fit the parameterization \cite{Chevallier:2000qy,Linder:2002et}
\be w(z)=w_0 + w_1 \frac{z}{1+z} \label{cplpar} \ee to current
data from various dark energy observational probes including the
Gold SnIa sample\cite{Riess:2004nr}, the first year SNLS
dataset\cite{Astier:2005qq}, the 3-year WMAP CMB shift
parameter\cite{Spergel:2006hy,Wang:2006ts}, the SDSS baryon
acoustic peak (BAO)\cite{Eisenstein:2005su}, the X-ray gas mass
fraction in clusters\cite{Allen:2002sr,Allen:2004cd} and the
linear growth rate of perturbations at $z=0.15$ as obtained from
the 2dF galaxy redshift survey\cite{Verde:2001sf,Hawkins:2002sg},
to investigate which of these probes currently favor crossing of
the PDL and what are the priors required. Finally in section IV we
conclude, summarize and outline future \nopagebreak prospects of
this work.

\section{Theoretical Implications of PDL Crossing}
The simplest class of physically motivated models generalizing the
cosmological constant and producing accelerated universe expansion
are based on a simple evolving homogeneous scalar field $\Phi$
(quintessence \cite{Caldwell:1997ii,Zlatev:1998tr,Ratra:1987rm})
minimally coupled to gravity whose dynamics is determined by a
potential $U(\Phi)$ and has a canonical kinetic term
$\frac{1}{2}{\dot \Phi}^2$. It is easy to see however that this
class of models can not reproduce a $w(z)$ crossing the PDL $w=-1$
for {\it any} potential $U(\Phi)$. Indeed, the equation of state $w$
for such models takes the form \be
w(z)=\frac{p_X}{\rho_X}=\frac{\frac{1}{2}{\dot \Phi}^2 -
U(\Phi)}{\frac{1}{2}{\dot \Phi}^2 + U(\Phi)} \label{wzmc} \ee which
approaches $w=-1$ in the limit of a small kinetic term but does not
cross the PDL as long as ${\dot \Phi}^2 >0$ for any sign of
$U(\Phi)$. This result has been generalized by
Vikman\cite{Vikman:2004dc} who showed that any minimally coupled
scalar field with a generalized kinetic term also can not cross the
PDL through a stable trajectory (see Refs
\cite{Cannata:2006gd,Andrianov:2005tm} for an interesting exception
involving however an arbitrary change of the kinetic term sign).

A simple `no go' theorem\cite{Amendola:2006pc,Kunz:2006wc} can
also be obtained for a general perfect barotropic fluid with
conserved energy momentum tensor whose equation of state is of the
form $p=p(\rho)$. Indeed energy conservation implies that \be
{\dot \rho} = -3\frac{\dot a}{a} (\rho + p(\rho)) \label{econ1}
\ee From this equation it follows that ${\dot \rho}\rightarrow 0$
as $p(\rho)\rightarrow -\rho$ or as $w=\frac{p}{\rho}\rightarrow
-1$. By differentiating equation (\ref{econ1}) with respect to
time it is easy to see that ${\ddot \rho}\rightarrow 0$ as
$w\rightarrow -1$ and similarly for all time derivatives of the
density $\rho$ ie \be \lim_{w\rightarrow
-1}\frac{d^n\rho(t)}{dt^n}=0 \label{limnd1} \ee for all $n$.
Therefore, any fluid with equation of state of the form
$p=p(\rho)$ can not cross the PDL. Instead it asymptotically
approaches a constant energy density $\rho$ thus mimicking a
cosmological constant. It should be stressed however that many
scalar field models do not behave like barotropic ($p=p(\rho)$)
fluids due to the dependence of $p$ and $\rho$ on two independent
variables ${\dot \Phi}$ and $\Phi$ and therefore this proof is not
equivalent to the proof that a minimally coupled scalar field can
not cross the PDL.

Since the minimal theoretical approaches are unable to reproduce a
PDL crossing we must turn to `non-minimal' models to achieve such
crossing. There are two main classes of such `non-minimal'
approaches: Either consider multiple component dark
energy\cite{Hu:2004kh,Caldwell:2005ai,Guo:2004fq,Kunz:2006wc,
Zhao:2006mp,Chimento:2006xu,Stefancic:2005sp,Wei:2006tn,Zhao:2005bu,Zhang:2006ck,Li:2005fm,Stefancic:2005cs}
(notice that higher derivative models and many non-barotropic fluids
are effectively multi-component) with at least one phantom degree of
freedom (eg scalar field with negative kinetic energy) or consider
extensions of general relativity
\cite{Boisseau:2000pr,Esposito-Farese:2000ij,Perivolaropoulos:2005yv,Perrotta:1999am,Aref'eva:2005fu,Apostolopoulos:2006si,
Amendola:1999er,McInnes:2005vp,Gumjudpai:2005ry,Onemli:2004mb,Carroll:2004hc,Nojiri:2005vv,Nojiri:2006je,
Sahni:2002dx,Alam:2005pb,Chimento:2006ac,Carroll:2004de,Nojiri:2006ri,Nojiri:2003ft}.
The phantom degrees of freedom of the former approach expressed by
phantom fields are plagued by catastrophic UV instabilities since
their energy is unbounded from below and allows vacuum decay through
the production of high energy real particles and negative energy
ghosts
\cite{Cline:2003gs,Buniy:2005vh,Dubovsky:2005xd,DeFelice:2006pg}. On
the other hand the general relativity extensions of the latter
approach are severely constrained by local solar system and by
cosmological observations but are well motivated theoretically.

To illustrate the multiple dark energy component approach to PDL
crossing with scalar fields let's consider a set of two coupled
real scalar fields (quintessence $\Phi_1$ + phantom $\Phi_2$ :
quintom dark energy \cite{Guo:2004fq}) with Lagrangian \be {\cal
L}= \frac{1}{2}(\partial_\mu \Phi_1)^2-\frac{1}{2}(\partial_\mu
\Phi_2)^2-U(\Phi_1,\Phi_2) \label{qlang} \ee The effective
pressure and energy density for a homogeneous system is \ba
p&=&\frac{1}{2} {\dot \Phi_1}^2 - \frac{1}{2} {\dot
\Phi_2}^2-U(\Phi_1,\Phi_2) \label{pquint} \\
\rho&=&\frac{1}{2} {\dot \Phi_1}^2 - \frac{1}{2} {\dot
\Phi_2}^2+U(\Phi_1,\Phi_2) \label{rquint} \ea leading to the
equation of state parameter \be w=\frac{{\dot \Phi_1}^2 -{\dot
\Phi_2}^2-2U(\Phi_1,\Phi_2)}{{\dot \Phi_1}^2 -{\dot
\Phi_2}^2+2U(\Phi_1,\Phi_2)} \label{wquint} \ee which crosses the
PDL line when ${\dot \Phi_1}^2 -{\dot \Phi_2}^2$ changes sign.
This can easily be achieved with appropriate potentials and
initial conditions.

The same approach may be illustrated by considering a mixture of
multiple perfect fluids
\cite{Hu:2004kh,Stefancic:2005sp,Caldwell:2005ai,Stefancic:2005cs,Nojiri:2005sr}
instead of scalar fields. Consider a mixture of two
non-interacting fluids $(\rho_1,p_1, w_1)$ and $(\rho_2,p_2, w_2)$
with separately conserved energies and constant equation of state
parameters $w_1$, $w_2$ with $w_1>-1$, $w_2<-1$. The equation of
state parameter $w$ of the mixture is \cite{Kunz:2006wc} \be
w=\frac{p}{\rho}=\frac{p_1 + p_2}{\rho_1 + \rho_2}=\frac{w_1
\rho_{01} a^{-3(1+w_1)}+w_2 \rho_{02} a^{-3(1+w_2)}}{\rho_{01}
a^{-3(1+w_1)}+ \rho_{02} a^{-3(1+w_2)}} \label{wmix} \ee which
interpolates between $w=w_1$ ($a<<1$) and $w=w_2$ ($a>>1$) thus
crossing the PDL $w=-1$.

The requirement of phantom degrees of freedom which are plagued with
several unattractive features
\cite{Abramo:2005be,Cline:2003gs,Buniy:2005vh,Dubovsky:2005xd,DeFelice:2006pg,Tsujikawa:2005aq}
(instabilities and lack of realistic prototypes) makes this class of
approaches theoretically unattractive compared to the second class
which is based on extensions of general relativity. Even though the
latter are severely constrained observationally they are strongly
motivated from the theoretical viewpoint. First it is clear that
general relativity is an incomplete theory because it does not
contain quantum mechanics and also it is plagued by singularities.
Second, all theories that attempt to quantize gravity and/or unify
it with other interactions require modifications of general
relativity. For example, the string theory dilaton
\cite{Gasperini:2002bn,Copeland:2006wr} field can be understood as
the scalar field of an effective scalar-tensor theory in
4-dimensions\cite{Levin:1994yw}. Similarly, all Kaluza-Klein type
theories which attempt to unify gravity with other interactions by
utilizing compactified extra dimensions are described as effective
scalar-tensor theories at low
energies\cite{Perivolaropoulos:2002pn,Perivolaropoulos:2003we}.

The simplest but very general (given its simplicity) extension of
general relativity is expressed through scalar-tensor theories. In
these theories Newton's constant obtains dynamical properties
expressed through the potential $F(\Phi)$. The dynamics are
determined by the Lagrangian density
\cite{Boisseau:2000pr,Esposito-Farese:2000ij} \be {\cal
L}=\frac{F(\Phi)}{2}~R - \frac{1}{2}~g^{\mu\nu}
\partial_{\mu}\Phi
\partial_{\nu}\Phi
- U(\Phi)  + {\cal L}_m[\psi_m; g_{\mu\nu}]\  \label{lst} \ee
where ${\cal L}_m[\psi_m; g_{\mu\nu}]$ represents matter fields
approximated by a pressureless perfect fluid. The function
$F(\Phi)$ is observationally constrained as follows:
\begin{itemize}
\item $F(\Phi)>0$ so that gravitons carry positive
energy\cite{Esposito-Farese:2000ij}. \item
$\frac{dF}{d\Phi}<10^{-5}$ from solar system observations
\cite{will-bounds}.
\end{itemize}
Assuming a homogeneous $\Phi$ and varying the action corresponding
to (\ref{lst}) in a background of a flat FRW metric, we find the
coupled system of generalized Friedman equations \ba 3F H^2 &=&
\rho + \frac{1}{2} {\dot \Phi}^2 -3H{\dot F} + U \label{stfe1} \\
-2F{\dot H} & = & \rho + p + {\dot \Phi}^2 + {\ddot F} - H{\dot F}
\label{stfe2} \ea where we have assumed the presence of a perfect
fluid $(\rho=\rho_m, p\simeq 0)$ playing the role of matter fields.
Expressing in terms of redshift and eliminating the potential $U$
from equations (\ref{stfe1}), (\ref{stfe2}) we find
\cite{Esposito-Farese:2000ij,Perivolaropoulos:2005yv} \ba
&&\Phi'^2=-F''-\left[(lnH)'+ \frac{2}{1+z}\right]F' + 2
\frac{(lnH)'}{1+z}F - \nn
\\&&-3(1+z)\omm \left(\frac{H_0}{H}\right)^2 F_0>0
\label{stphp} \ea

\noindent where $'$ denotes derivative with respect to redshift and
$F_0$ is set to 1 in units of $\frac{1}{8\pi G_N}$ and corresponds
to the present value of $F$. In the limit of minimal coupling
($F=1$) equation (\ref{stphp}) becomes \be \Phi_{mc}^{'2}(H)=  2
\frac{(lnH)'}{1+z}-3(1+z)\omm (\frac{H_0}{H})^2>0 \label{mcphp} \ee
Using equation (\ref{wzh1}) it is easy to show that the inequality
(\ref{mcphp}) is equivalent to $w(z)>-1$ which confirms the
expectation that a minimally coupled scalar field is not consistent
with PDL crossing. This constraint however is relaxed due to the
allowed dynamical evolution of $F$ in equation (\ref{stphp}) which
may be written as \be \Phi^{'2}(F,H)=\Phi_{mc}^{'2}(H)-\Delta(F,H)>0
\label{stmcpp} \ee where \be \Delta(F,H)=F''+\left[(lnH)'+
\frac{2}{1+z}\right]F' + 2 \frac{(lnH)'}{1+z}(1-F) \label{dfh} \ee
Therefore, the constraint $\Phi_{mc}^{'2} (H)>0$ is replaced in
scalar-tensor theories by $\Phi_{mc}^{'2} (H)>\Delta(F,H)$ and by
choosing an $F$ such that $\Delta(F,H)<0$ a PDL crossing can be
achieved.

The question to be addressed is the following: `Do observational
constraints on $F$ allow $\Delta(F,H)<0$?' The answer to that
question is positive as may be easily seen by considering the
present time ($z=0$) when the observational constraints
\cite{Umezu:2005ee} (and especially the solar system
tests\cite{will-bounds}) are most stringent requiring $F'(z) \simeq
0$ but placing no constraints on $F''(z=0)$. Setting $z=0$,
$F=F_0=1$, $F'=0$ in equation (\ref{stphp}) we obtain \ba
\left(\frac{H^2}{H_0^2}\right)'(z&=&0)>3\omm + F''(z=0)\nn \\
\implies w(z&=&0)+1>\frac{F''(z=0)}{3(1-\omm)}\label{stcon} \ea
where we have used equation (\ref{wzh1}). Since there is no
observational constraint on $F''(z=0)$ we can clearly pick
$F''(z=0)<0$ allowing for $w(z=0)<-1$ and crossing the PDL.

There is a simple physical interpretation of the behavior required
by $F$ in order to cross the PDL and lead to superacceleration
($w<-1$). Equation (\ref{dfh}) implies that $\Delta(H,F)<0$ (and
therefore $w<-1$) is favored for $F''<0$, $F'<0$ and $F>1$. Since
$F\sim G_{eff}^{-1}$ this behavior implies an effective Newton's
constant that increases with redshift ($G''_{eff}>0$,
$G'_{eff}>0$) and therefore decreases with time (${\ddot
G}_{eff}<0$, ${\dot G}_{eff}<0$, $G_{eff}(t)>G_{eff}(t_0)$) thus
`helping' the accelerating expansion induced by the potential $U$.
This type of behavior was verified in a specific reconstruction
example of $F(z)$, $U(z)$ in Ref. \cite{Perivolaropoulos:2005yv}
corresponding to an $H(z)-w(z)$ that crosses the PDL. It should be
stressed however that any ansatz of $G_{eff}$ that is monotonic
function of time, eg $G_{eff}(t) \sim t^{-\alpha}$ where
$\alpha>0$, can lead to very tight constraints on the present
values of $G'_{eff}$ and $G''_{eff}$ using nucleosynthesis
constraints\cite{Copi:2003xd} (but see also \cite{DeFelice:2005bx}
for a way to relax such constraints).

To summarize we have discussed two broad classes of models which can
lead to crossing of the PDL: multicomponent dark energy with phantom
components and extensions of general relativity. This is not an
exhaustive classification but many of the left out models can be
incorporated in the above classes. For example models of coupled
quintessence where matter density \cite{Amendola:1999er} is
explicitly coupled to the scalar field causing acceleration may be
shown to be conformally equivalent to scalar-tensor theories.
Similarly, braneworld models
\cite{Sahni:2002dx,Apostolopoulos:2006si,Aref'eva:2005fu,Alam:2005pb,
Chimento:2006ac,Bogdanos:2006pf,Kofinas:2005hc} can be classified as
extended gravity theories. On the other hand in the multi-component
class we could include models with a complex equation of
state\cite{Arbey:2006it,Kunz:2006wc}, fermion models, scalar fields
with higher derivatives\cite{Li:2005fm} and vector field
models\cite{Zhao:2005bu}. Of the two classes of theories discussed,
the class of extended gravity theories is strongly motivated
theoretically in contrast to the multicomponent class. Therefore,
despite the observational constraints, extensions of general
relativity is the prime candidate class of theories consistent with
PDL crossing.

It is thus important to address the following questions:
\begin{itemize}
\item Do current cosmological data support a crossing of the PDL?
\item What are the optimal observational strategies to confirm or
exclude a PDL crossing with future observations?
\end{itemize}
The main goal of the next section is to address the first question
using a broad sample of cosmological data. Previous studies
\cite{Nesseris:2004wj} have indicated that the SnIa data do not show
a common trend regarding the crossing of the PDL. In particular, the
analyses of the Gold SnIa dataset have indicated that a $w(z)$ that
crosses the PDL is preferred over the $\Lambda$CDM parameter values
at a level of almost $2\sigma$
\cite{Alam:2004jy,Wang:2003gz,Lazkoz:2005sp,alam1}. On the other
hand the SNLS dataset does not show such a trend
\cite{Nesseris:2005ur,Huterer:2004ch,Santos:2006ja,Jassal:2006gf,Xia:2005ge}
and the $\Lambda$CDM parameter values are well within $1\sigma$ from
the best fit parameter values. This raises the question as to what
is the trend favored by the other than SnIa cosmological data. Do
they favor the trend of PDL crossing indicated by the Gold dataset
or do they favor a constant $w\simeq -1$ as indicated by SNLS? The
main goal of the next section will be to address this question using
the following cosmological data:
\begin{itemize}
\item The location of the CMB perturbation spectrum
peaks\cite{Spergel:2006hy} (shift
parameter\cite{Bond:1997wr,Trotta:2004qj,Melchiorri:2000px}).
\item The measurement of the CMB acoustic scale at $z_{BAO}=0.35$ as
indicated by the large scale structure correlation function of SDSS
redshift survey (Baryon Acoustic Oscillations (BAO) peak)
\cite{Eisenstein:2005su}. \item The gas mass fraction of galaxy
clusters $f_{gas}=\frac{M_{gas}}{M_{tot}}$ inferred from X-ray
observations \cite{Allen:2004cd,Allen:2002sr}, which depends on the
angular diameter distance $d_A$ to the cluster as $f_{gas}\sim
d_A^{3/2}$. \item The linear growth factor $D(a)$ (eq. \ref{dadef})
as determined by the 2dF galaxy redshift
survey\cite{Verde:2001sf,Hawkins:2002sg} which can constrain $H(z)$
by using eq. (\ref{greq}) in the context of general relativity.
\end{itemize} Using the above cosmological data we find the best
fit form of $w(z)$ with $1\sigma$ errors in the context of the CPL
parameterization\cite{Chevallier:2000qy,Linder:2002et} \ba w(a) &=&
w_0 + w_1 (1-a)  \\ w(z) &=& w_0 + w_1 \frac{z}{1+z} \label{cplpar}
\ea We then compare this form with the corresponding best fits of
both the Gold and the SNLS datasets to see which of the two trends
is favored. The comparison is made for priors of $\Omega_{0m}$ in
the range $0.2-0.3$.

\section{Observational  of PDL Crossing}
\subsection{The SnIa datasets}
The two most reliable and robust SnIa datasets existing at present
are the Gold dataset \cite{Riess:2004nr} and the Supernova Legacy
Survey (SNLS) \cite{Astier:2005qq} dataset. The Gold dataset
compiled by Riess et. al. is a set of supernova data from various
sources analyzed in a consistent and robust manner with reduced
calibration errors arising from systematics. It contains 143
points from previously published data plus 14 points with $z>1$
discovered recently with the HST. The SNLS is a 5-year survey of
SnIa with $z<1$. It has recently \cite{Astier:2005qq} released the
first year dataset. The SNLS has adopted a more efficient SnIa
search strategy involving a `rolling search' mode where a given
field is observed every third or fourth night using a single
imaging instrument thus reducing photometric systematic
uncertainties. The published first year SNLS dataset consists of
44 previously published nearby SnIa with $0.015<z<0.125$ plus 73
distant SnIa ($0.15<z<1$) discovered by SNLS two of which are
outliers and are not used in the analysis. %%%%
The fact that in the two datasets a set of low-z SnIa is common to
both samples could only lead to minor common systematics due to
low redshift.

The above observations provide the apparent magnitude $m(z)$ of
the supernovae at peak brightness after implementing correction
for galactic extinction, K-correction and light curve
width-luminosity correction. The resulting apparent magnitude
$m(z)$ is related to the luminosity distance $D_L(z)$ through \be
m_{th}(z)={\bar M} (M,H_0) + 5 log_{10} (D_L (z)) \label{mdl} \ee
where in a flat cosmological model \be D_L (z)= (1+z) \int_0^z
dz'\frac{H_0}{H(z';a_1,...,a_n)} \label{dlth1} \ee is the Hubble
free luminosity distance ($H_0 d_L$), $a_1,...,a_n$ are
theoretical model parameters and ${\bar M}$ is the magnitude zero
point offset and depends on the absolute magnitude $M$ and on the
present Hubble parameter $H_0$ as \ba
{\bar M} &=& M + 5 log_{10}(\frac{H_0^{-1}}{Mpc}) + 25= \nn \\
&=& M-5log_{10}h+42.38 \label{barm} \ea The parameter $M$ is the
absolute magnitude which is assumed to be constant after the above
mentioned corrections have been implemented in $m(z)$.

The data points of the Gold dataset are given after the
corrections have been implemented, in terms of the distance
modulus \be \mu^G_{obs}(z_i)\equiv m^G_{obs}(z_i) - M
\label{mug}\ee The SNLS dataset however also presents for each
point, the stretch factor $s$ used to calibrate the absolute
magnitude and the rest frame color parameter $c$ which mainly
measures host galaxy extinction by dust. Thus, the distance
modulus in this case depends apart from the absolute magnitude
$M$, on two additional parameters $\alpha$ and $\beta$ defined
from \be
\mu_{obs}^{SNLS}=m_{obs}^{SNLS}(z_i)-M+\alpha(s_i-1)-\beta c_i
\label{musnls} \ee which are fit along with the theoretical
parameters using a recursive procedure.

\begin{figure}[t]
\hspace{0pt}\rotatebox{0}{\resizebox{0.50\textwidth}{!}{\includegraphics{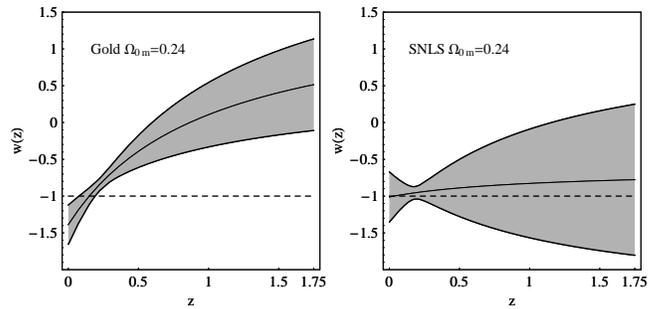}}}
\vspace{-10pt}{\caption{The best fit form of $w(z)$ for the Gold
and SNLS datasets for a prior of $ \Omega_{0m} = 0.24 $ along with
the $1\sigma$ errors (shaded region).}} \label{fig2}
\end{figure}

The theoretical model parameters are determined by minimizing the
quantity \be \chi^2_{SnIa} (\omm,w_0,w_1)= \sum_{i=1}^N
\frac{(\mu_{obs}(z_i) - \mu_{th}(z_i))^2}{\sigma_{\mu \; i}^2 +
\sigma_{int}^2 + \sigma_{v\; i}^2 } \label{chi2} \ee where $N=115$
for SNLS ($\chi^2_{SNLS}$), $N=157$ for the Gold dataset
($\chi^2_{Gold}$), $\sigma_{\mu \; i}^2$, $\sigma_{int}^2$ and
$\sigma_{v\; i}^2$ are the errors due to flux uncertainties,
intrinsic dispersion of SnIa absolute magnitude and peculiar
velocity dispersion respectively. These errors are assumed to be
gaussian and uncorrelated. The theoretical distance modulus is
defined as \be \mu_{th}(z_i)\equiv m_{th}(z_i) - M =5 log_{10}
(D_L (z)) +\mu_0 \label{mth} \ee where \be \mu_0= 42.38 - 5
log_{10}h \label{mu0}\ee and $\mu_{obs}$ is given by (\ref{mug})
and (\ref{musnls}) for the Gold and SNLS datasets respectively.

The steps we followed for the minimization of (\ref{chi2}) for the
Gold and SNLS datasets are described in detail in Refs
\cite{Nesseris:2004wj,Nesseris:2005ur,Nesseris:2006jc}. The form
of $H(z)$ used in (\ref{dlth1}), (\ref{mth}) is obtained by using
the CPL parameterization (\ref{cplpar}) and leads to \ba H^2
(z)&=&H_0^2 [ \Omega_{0m} (1+z)^3 + \nn
\\ &+& (1-\Omega_{0m})(1+z)^{3(1+w_0+w_1)}e^{\frac{-3w_1 z}{(1+z)}}]
\label{hcpl} \ea The best fit form of $w(z)$ for the Gold and first
year SNLS datasets is shown in Fig. 2 for a prior of $\omm=0.24$
along with the $1\sigma$ errors\cite{numerics} (shaded region). Even
though the two fits are consistent at the $1\sigma$ level it is
clear that the Gold sample mildly favors a crossing of the PDL at
$z\simeq 0.25$ while no such trend appears for the SNLS data. This
trend difference could be attributed to three factors
\begin{itemize}
\item Statistical errors \item Systematic errors \item Different
redshift ranges (Gold $0<z<1.7$, SNLS $0<z<1$). \end{itemize} The
third possibility has been excluded in Ref. \cite{Nesseris:2005ur}
where a truncated version of the Gold sample with $(0<z<1)$ was
found to show even stronger trend for crossing the PDL than the full
Gold sample. Given this apparent difference in trends between the
Gold and the SNLS datasets the following question arises: `Which (if
any) of the two trends do other cosmological probes of dark energy
favor?'

\subsection{The CMB shift parameter}
A particularly accurate and deep geometrical probe of dark energy is
the angular scale of the sound horizon at the last scattering
surface as encoded in the location $l_1^{TT}$ of the first peak of
the CMB temperature perturbation spectrum. By measuring the angular
scale $\theta_1^{TT}\sim 1/l_1^{TT}$ of the last scattering sound
horizon and calculating its comoving scale $r_s(z_{dec})$
independently from CMB physics, the sound horizon angular diameter
distance $d_A(z_{rec})$ can be obtained assuming flatness as \be
d_A(z_{rec})=\frac{r_s(z_{rec})}{
\theta_1^{TT}}=\frac{1}{1+z_{rec}}\int_0^{z_{rec}}\frac{dz}{H(z)}
\label{dazrec} \ee thus providing a useful constraint on $H(z)$. To
eliminate the model dependence involved in the calculation of the
sound horizon scale an (approximately) model independent parameter
can be defined by dividing the measured angle $\theta_1^{TT}\sim
1/l_1^{TT}$ by the corresponding angle $\theta_1^{'TT}$ of a
reference model. This parameter is known as the {\it `shift
parameter'}.

The shift parameter is defined as
\cite{Bond:1997wr,Efstathiou:1998xx,Trotta:2004qj,Melchiorri:2000px}
\be R=\frac{l_1^{'TT}}{l_1^{TT}} \label{shpardef}\ee where
$l_1^{TT}$ is the temperature perturbation CMB spectrum multipole of
the first acoustic peak. In the definition of $R$, $l_1^{TT}$
corresponds to the model (with fixed $\Omega_{0m}$, $\Omega_{0b}$
and $h$) characterized by the shift parameter and $l_1^{'TT}$ a {\it
reference} flat SCDM model ($\Omega'_{0m}=1$) with the same
$\omega_m\equiv\Omega_{0m}h^2$, $\omega_b=\Omega_{0b}h^2$ as the
original model.

The location $l_1^{TT}$ of the first acoustic peak can be connected
with the angular diameter distance $d_A$ to the last scattering
surface and with the sound horizon $r_s$ at the last scattering
surface ($z=z_{rec}$) as follows \cite{Efstathiou:1998xx,Hu:2000ti}:
\be l_1^{TT}=l_A(d_A,r_s) (1-\psi(\omega_m,\omega_b)) \label{l1eq}
\ee
where \ba l_A(d_A,r_s) &\equiv& \frac{\pi}{\theta_1^{TT}} \\
\theta_1^{TT} &\equiv& \frac{r_s(z_{rec})}{d_A(z_{rec})} \ea and the
phase shift parameter $\psi(\omega_m,\omega_b)\simeq 0.27$
\cite{Hu:2000ti} depends weakly on cosmological parameters. The
sound horizon $r_s$ and the angular diameter distance $d_A$ depend
on the Hubble expansion history (at early and late times
respectively) as follows: \ba r_s&=&a_{rec} \int_0^{a_{rec}}
\frac{c_s(a) da}{a^2 H(a)}= \nn \\ &=& a_{rec} \int_0^{a_{rec}}
\frac{c_s(a) da}{\omm^{1/2}}\left[ \frac{\omr h^2}{\omm h^2} +
a\right]^{-1/2}\label{shor}\ea where $c_s(a)$ is the sound velocity
which at decoupling ($a=a_{rec}$) is \be c_s^2(a_{rec})=\frac{\delta
p}{\delta \rho} = \frac{1}{3}\frac{1}{1+\frac{3\omb}{4\omr}a_{rec}}
\ee and \be d_A(z)=\frac{a_{rec}}{H_0 \sqrt{\omk}}sin\left[H_0
\sqrt{\omk}\int_0^{z_{rec}} \frac{dz'}{H(z')}\right] \ee

\begin{figure}
\hspace{0pt}\rotatebox{0}{\resizebox{0.5\textwidth}{!}{\includegraphics{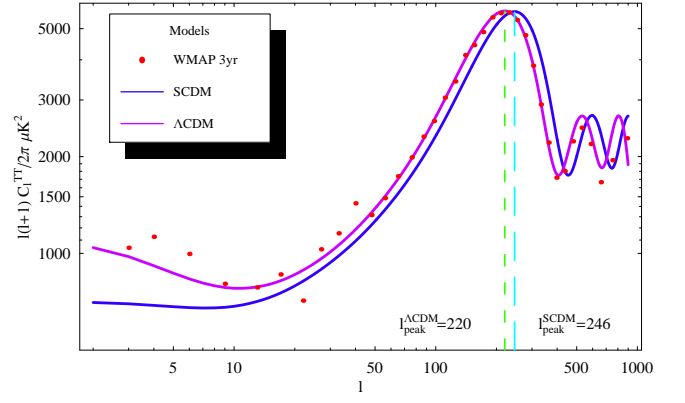}}}
\vspace{-10pt}{\caption{{\it Derivation of the shift parameter:} The
3-year WMAP binned data along with a theoretical flat $\Lambda$CDM
model ($\omega_m =0.14, \omega_b =0.022,h=0.72 $) and a SCDM model
with ($\omega_m =0.14, \omega_b =0.022,h=0.37 $). For the
$\Lambda$CDM the first peak is at $l_1^{TT}=220$ while for the SCDM
model at $l_1^{'TT}=246$.}} \label{fig3}
\end{figure}

For the reference model SCDM we have \ba
d'_A(z'_{rec})&=&\frac{a_{rec}}{H_0}\int_0^{z_{rec}}
\frac{dz'}{\left[\omm' (1+z)^3 +\omr' (1+z)^4\right]^{1/2}}=\nn \\
&=&
\frac{2a_{rec}}{H_0}\left[(a'_{eq}+1)^{1/2}-(a'_{{rec}}+a'_{eq})^{1/2}\right]=\nn\\
&=&\frac{2
a_{rec}}{H_0}\left[(\omr'+1)^{1/2}-(a'_{{rec}}+\omr')^{1/2}\right]\label{daz1}
\\ &\equiv& \frac{2a_{rec}}{H_0} q(\omr',a'_{rec})\nn \ea Using equations (\ref{l1eq})-(\ref{daz1}) it is easy to show that \ba
R&=&\frac{l_1^{'TT}}{l_1^{TT}}=\frac{l_A^{'}(1-\psi')}{l_A(1-\psi)}=\nn
\\ &=& \frac{r_s}{r'_s}\frac{d'_A (z'_{rec})}{d_A
(z_{rec})}=\frac{2}{\omm^{1/2}}\frac{q(\omr',a_{rec})}{\chi(z)}\label{shparnew}
\ea where \be \chi(z)\equiv \int_0^{z}\frac{H_0 dz'}{H(z')} \ee and
we have assumed flatness and used the fact that $\psi=\psi'$ (since
$\omega_b=\omega'_b$ and $\omega_m=\omega_m'$) and
$a_{rec}(\omega_b,\omega_m)=a'_{rec}(\omega'_b,\omega'_m)$
\cite{Hu:2000ti}. The main advantage of using the shift parameter
instead of just the value of $l_1^{TT}$ is that it is weakly
dependent on parameters other than geometry ($H(z)$). The dependence
on other parameters enters through $\omm^{-1/2}$ and through
$q(\omr',a_{rec})=q(\omega_r/h'^2,a_{rec}(\omega_m,\omega_b))$ where
$\omega_r\equiv \omr h^2\simeq 2.47 \cdot 10^{-5}$, $h'^2=\omega_m$
and  \cite{Hu:2000ti} $z_{rec}=\frac{1}{a_{rec}}-1=1048 (1+0.00124
\omega_b^{-0.738})(1+g_1 \omega_m^{g_2})$ where $g_1=0.0783
\omega_b^{-0.238}(1+39.5 \omega_b^{0.763})^{-1}$ and $g_2=0.560
(1+21.1 \omega_b^{1.81})^{-1}$. Notice that the weak dependence of
the shift parameter on $\omega_m$, $\omega_b$ through
$q(\omega_m,\omega_b)$ has not been explicitly demonstrated in
previous studies \cite{Wang:2006ts}. Instead, the shift parameter is
usually expressed as \be
R'=\frac{2}{\omm^{1/2}\chi(z_{rec})}\label{shparold} \ee or
equivalently as $ {\bar R}=\omm^{1/2}\chi(z_{rec})$ thus omitting
the correction factor $q(\omega_m,\omega_b)$. For fixed $\omm=0.27$,
$\omb=0.043$ and $0.5\leq h \leq 0.9$, $q(\omega_m,\omega_b)$
remains in the range $0.964\leq q \leq 0.968$ and therefore the
dependence on the Hubble parameter introduced by $q$ is very weak.

The main disadvantage of the shift parameter is that in order to
evaluate it using its definition (\ref{shpardef}) for a given CMB
perturbation spectrum we need not only the location of the first
peak $l_1^{TT}$ but also the location of the reference model (SCDM)
peak. For an accurate result, the latter requires running a
Boltzmann code like CAMB \cite{Lewis:1999bs} with $\omm'=1$,
$\omega'_m=\omega_m$, $\omega'_b=\omega_b$ and evaluating the
location of the first peak $l_1^{'TT}$. This procedure is
illustrated in Fig. 3 where we show the 3-year WMAP data along with
a theoretical fit obtained for
$(\omega_m,\omega_b,h)=(0.14,0.022,0.72)$ assuming flat
$\Lambda$CDM. The first peak for the best fit model is obtained at
$l_1^{TT}=220\pm 0.8$ while for the reference flat SCDM model with
$(\omega_m,\omega_b,h)=(0.14,0.022,0.37)$ (leading to $\omm=1$) we
have $l_1^{'TT}=246$ (see Fig. 3). We thus obtain \be
R=\frac{l_1^{'TT}}{l_1^{TT}}=1.123\pm 0.03 \label{rnum1} \ee Using
now (\ref{shparnew}) and (\ref{shparold}) with $q=0.965$ we find \be
{\bar R}\equiv \omm^{1/2} \chi(z_{rec})=1.71\pm 0.05 \label{rbnum}
\ee This result is consistent but slightly different from that of
Ref. \cite{Wang:2006ts} which used MCMC chains to obtain ${\bar R}$
from WMAP3 as \be {\bar R}=1.70\pm 0.03 \label{wshv}\ee In what
follows we adopt the result of eq. (\ref{wshv}) to obtain the best
fit $w(z)$.

Even though the shift parameter incorporates most of the geometrical
information of the CMB spectrum the following question can be
raised: `How much of the CMB spectrum geometrical information is
left out of the shift parameter?' To answer this question we proceed
in two steps: First we consider the CPL parameterization and
construct CMB spectra (using a modified version of
CAMB\cite{Lewis:1999bs}) varying $H(z)$ with fixed ${\bar R}$ and
other CMB parameters (see Fig. 4 and Table \ref{tablemodels}). The
constructed CMB spectra are practically identical for all values of
$l$ except low $l$ ($l\lsim 50$) where there are minor differences
due to the ISW effect \cite{Pogosian:2005ez,Melchiorri:2002ux}.
These CMB spectra differences can not be observationally
distinguished due to the large cosmic variance errors that dominate
at low $l$.

The temperature anisotropy due to the ISW effect is given by an
integral over the time variation of the gravitational potential
traversed by the photons coming from the last scattering
surface\cite{Pogosian:2005ez} \be \Theta_l(k,\eta_0)=(2l+1)
\int_{\eta_{rec}}^{\eta_0}d\eta e^{-\tau}\left[ 2{\dot
\Phi}(k,\eta)\right]j_l(k(\eta_0-\eta)) \label{iswth1} \ee where the
overdot indicates conformal time derivative, $\Theta\equiv
\frac{\Delta T}{T}$ and the integral is taken over the line to the
last scattering surface. Due to the integral over the oscillating
Bessel functions $j_l$, the dominant contribution to the CMB
anisotropy is on large scales (low $l$). The gravitational potential
$\Phi(k,\eta)$  (metric fluctuation of FRW background) may be shown
(using the Einstein equations) to

\begin{figure}
\rotatebox{0}{\hspace{-10pt}\resizebox{0.53\textwidth}{!}{\includegraphics{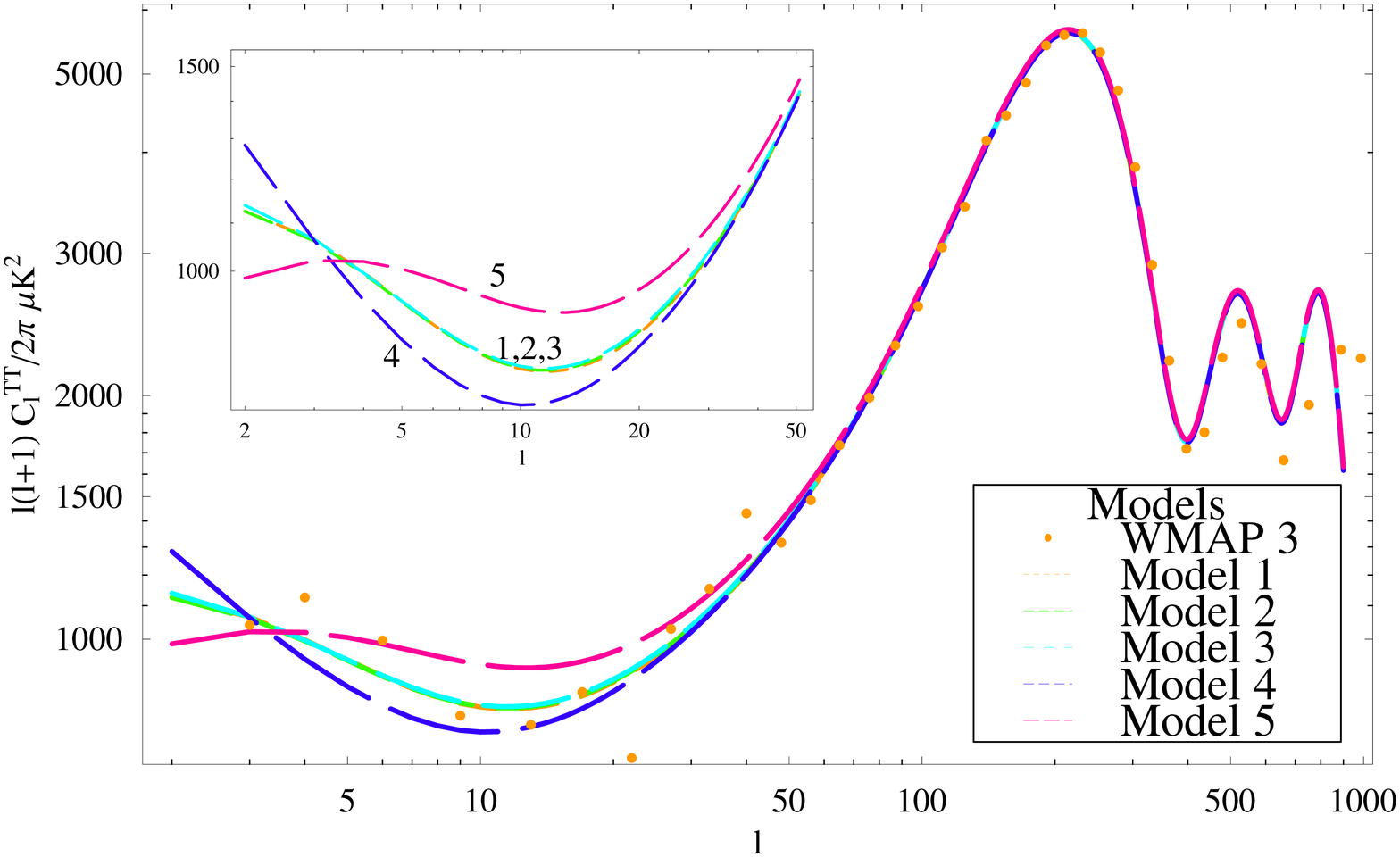}}}
\vspace{-7pt}{\caption{Theoretical CMB spectra for the CPL
parameterization for five models that have fixed ${\bar R}=1.70$ and
other parameters (see text and Table \ref{tablemodels}). Models
(1,2,3) with identical $\omm$ and ${\bar R}$ but different $w(z)$
have practically identical spectrum for all $l$ (identical ISW
effect). The ISW effect for fixed ${\bar R}$ changes only due to
variations of $\omm$ (models 4, 5)!}} \label{fig4}
\end{figure}

\noindent obey a Poisson equation which in a flat universe is of
the form \be k^2 \Phi(k, \eta)=4\pi G a^2 \delta \rho
\label{pois1} \ee where $\delta \rho$ is the density perturbation.
In a flat, matter dominated universe we have $\frac{\delta
\rho}{\rho} \sim a \implies \delta \rho \sim a^{-2} $ which
implies (due to (\ref{pois1})) that ${\dot \Phi}=0$ and there is
no ISW effect. However, in the presence of dark energy, the
evolution of perturbations is delayed at late times
$\left(\frac{\delta \rho}{\rho}< \left(\frac{\delta
\rho}{\rho}\right)_i a\right)$ and therefore ${\dot \Phi}\neq 0$
leading to a non-zero ISW effect. The ISW effect therefore
measures the growth rate of perturbations at late times or its
deviation from the flat matter dominated rate $\frac{\delta
\rho}{\rho}\sim a$.

\begin{table}[b]
\caption{\label{tablemodels}Parameters for the 5 models of Fig.
4.}
\begin{center}
\begin{tabular}{lllll}
Models \;\;\;\;\;\; & $\Omega_{0m}$ \;\;\;\;\;\; & $w_0$ \;\;\;\;\;\; & $w_1$ \;\;\;\;\;\; & $h$\;\;\;\;\;\;\\
\hline
Model 1 &0.27 & -0.8 &-0.123  &0.725\\
Model 2 &0.27 & -0.9 &\;0.196 &0.725\\
Model 3 &0.27 & -1.0 &\;0.497 &0.725\\
Model 4 &0.14 & -1.7 &-1.061  &0.998\\
Model 5 &0.50 & -0.3 &-0.064  &0.533\\
\hline
\end{tabular}
\end{center}
\end{table}

An interesting feature of Fig. 4 is that if we keep ${\bar R}$ and
$\omm$ fixed while varying $H(z)$, the ISW effect appears to be
unchanged! In contrast, if we keep ${\bar R}$ fixed while varying
$\omm$ and $H(z)$ ($\rho_X(z)$), the ISW effect changes indicating
that perhaps there is not much more information in the ISW effect
about the dark energy density evolution than that included in the
shift parameter ${\bar R}$. This is also demonstrated in Fig. 5
where we show the growth of perturbations $D(a)$ for the five models
of Fig. 4. Clearly, those models which have identical ISW effect
\begin{figure}
\hspace{0pt}\rotatebox{0}{\resizebox{0.5\textwidth}{!}{\includegraphics{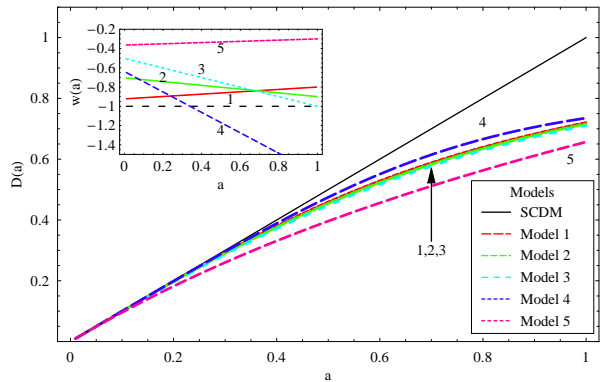}}}
\vspace{-10pt}{\caption{The growth of perturbations $D(a)$ for the
five models of Fig. 4 along with the corresponding forms of $w(a)$.
Those models (1,2,3) which have identical ISW effect (same $\omm$
and ${\bar R}$ but different $H(z)-w(z)$) also have very similar
growth of perturbations even though the dark energy density
evolution is quite different.}} \label{fig5}
\end{figure}
\noindent (same $\omm$ and ${\bar R}$ but different $H(z)$) also
have very similar growth of perturbations even though the dark
energy density evolution is quite different!

It should be stressed that whatever information about dark energy
is encoded in the ISW effect, this information can not be
extracted directly from the temperature perturbation CMB spectrum
(due to the large cosmic variance errors at low $l$) but only
using alternative methods like CMB cross
correlation\cite{Pogosian:2005ez} with large scale structure or
CMB polarization\cite{Cooray:2002cb}.

The measurement of the shift parameter (equation (\ref{rbnum}))
allows us to add an important term to the $\chi^2$ of equation
(\ref{chi2}) obtained with SnIa. This term is of the form \be
\chi^2_{CMB}(\omm,w_0,w_1)=\frac{({\bar
R}(\omm,w_0,w_1)-1.70)^2}{0.03^2} \label{chi2cmb} \ee This term is
very important in constraining evolving dark energy models for two
reasons: \begin{enumerate}
\item The shift parameter has a very low relative error (few
percent) \item The integral over $H(z)$ extends to high redshift
($z_{rec}\simeq 1089$). Therefore, minor modifications of $H(z)$
can be very `costly' in $\chi^2$ exactly due to this term.
\end{enumerate}

Another similar term is obtained from the imprint of the last
scattering sound horizon on the large scale structure correlation
function.

\subsection{The Baryon Acoustic Peak}

The large scale correlation function measured from the luminous
red galaxies spectroscopic sample of the SDSS (Sloan Digital Sky
Survey) \cite{Eisenstein:2005su} includes a clear peak at about
$100h^{-1}Mpc$. This peak was identified with the expanding
spherical wave of baryonic perturbations originating from acoustic
oscillations at recombination. The comoving scale of this shell at
recombination is about 150Mpc in radius. The identification of the
comoving scale where the correlation peak was observed with the
acoustic horizon at recombination requires the appropriate form of
the cosmological model $H(z)$ in converting from the observed
angles-redshift to correlation function distances. An accurate
determination of the best fit $H(z)$ would therefore correspond to
using a general model or parameterization of $H(z)$ to convert
from the observed angles-redshifts to correlation function
distances and then vary these parameters until the observed Baryon
Acoustic Oscillation (BAO) peak coincides with the expected value
from CMB physics.

A simplified version of this approach was used in Refs.
\cite{Eisenstein:2005su}, \cite{Blake:2003rh}. It involves the use
of a fiducial $\Lambda$CDM $H(z)$ model to construct the correlation
function from the observed angles-redshifts and identify the
distance scale $r_{peak}$ where the BAO peak appears. Comparing this
value of $r_{peak}$ with the expected value $r'_{peak}$ from CMB
physics would require a scale shift by a factor $\alpha$:
$r'_{peak}=\alpha r_{peak}$. This factor $\alpha$ can be
approximated as the ratio of the required characteristic distance
scale $D_V (z_{BAO})$ of the survey with mean redshift $z_{BAO}$
(obtained with the correct $H(z)$) over the distance scale
$D_V^{\Lambda CDM}(z_{BAO})$ corresponding to the fiducial
$\Lambda$CDM model \be \alpha=\frac{r'_{peak}}{r_{peak}}=\frac{D_V
(z_{BAO})}{D_V^{\Lambda CDM}(z_{BAO})} \label{dvdvl} \ee

The characteristic distance scale of the redshift survey with
typical redshift $z_s$ can be connected to $H(z)$ as follows:
Consider a spherical shell (Fig. 6) of comoving radius R. Let $z$
be the redshift corresponding to the sphere (points C and D) as
viewed by an observer at O and $\Delta z$ the redshift difference
between A and B.

\begin{figure}
\rotatebox{270}{\hspace{0pt}\resizebox{.8\textwidth}{!}{\includegraphics{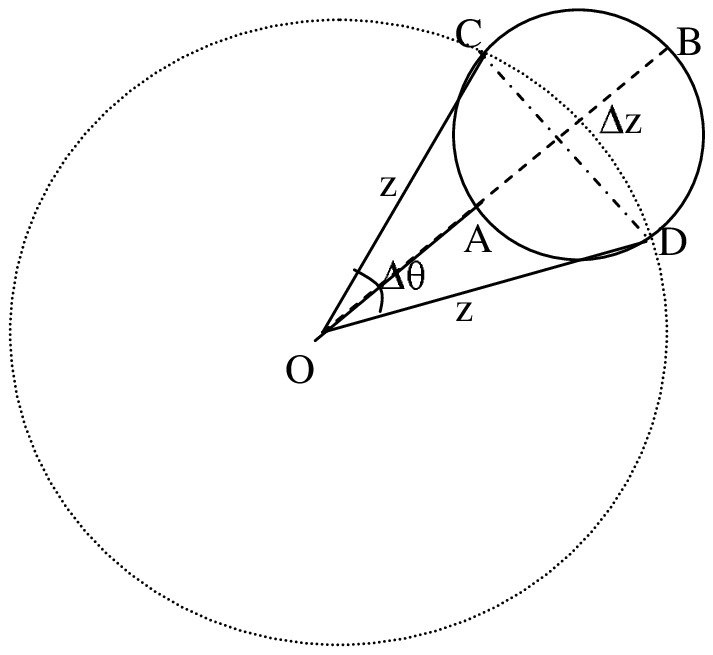}}}
\vspace{-250pt}\caption{A spherical shell in redshift space.}
\label{fig6}
\end{figure}

If we measure the angular scale $\Delta \theta$, the redshifts $z$
and $\Delta z$, then given a cosmological model $H(z)$, the comoving
scales CD and AB may be evaluated from the flat FRW metric
as\cite{Alcock:1979mp} \be CD= x \Delta \theta =\Delta
\theta\int_{t_i}^{t_0} \frac{dt}{a}=\frac{\Delta
\theta}{H_0}\int_0^z \frac{dz}{E(z)} \ee where $E(z)\equiv H(z)/H_0$
and \be AB=\Delta x=\frac{\Delta t}{a}=\frac{1}{H_0} \frac{\Delta
z}{E(z)} \ee Generalizing the above to a deformed sphere ($AB\neq
CD$) we may associate an approximate single scale to the sphere as
\cite{Eisenstein:2005su} \ba D_V (z)&=&\left[\left(\frac{CD}{\Delta
\theta}\right)^2(AB)\right]^{1/3}=\nn
\\&=&\left[\left(\frac{1}{H_0}\int_0^z
\frac{dz}{E(z)}\right)^2\frac{1}{H_0} \frac{\Delta
z}{E(z)}\right]^{1/3} \ea where $ CD/ \Delta \theta $ is the
comoving angular diameter distance. We also want this scale to be
representative of the whole sample (fiducial scale for SDSS
$z_{BAO}=0.35$) so we allow the sphere to extend from the observer
to ($A\rightarrow O$) to $z\simeq 0.35$ ($z_{BAO} \simeq 0.35$) and
therefore $\Delta z\simeq z_{BAO}\simeq 0.35$. For low redshift
samples (like $z_{BAO}=0.35$), the dilation scale $D_V(z)$ can be
assumed to encode all the information required for converting
redshift to distance. By demanding that the scale of the peak of the
correlation function coincides with the acoustic scale (which is
predicted by simple physics (see eq. (\ref{shor})) and depends on
$\omm h^2$) it has been shown \cite{Eisenstein:2005su} that \be D_V
(0.35)=1370\pm 64 Mpc \ee for $\omm h^2= 0.130\pm 0.010 $. A
dimensionless and independent of $H_0$ version of the dilation scale
$D_V$ is \ba A &\equiv & D_V(0.35) \frac{\sqrt{\omm
H_0^2}}{z_{BAO}}=\nn
\\ &=& \nn \omm^{1/2}
E(z_{BAO})^{-1/3}[\frac{1}{z_{BAO}}\int_0^{z_1}\frac{dz'}{E(z')}]^{2/3}=
\\ &=&0.469 \pm 0.017 \label{ameas1} \ea  This parameter has been
used extensively in constraining dark energy models (see eg
\cite{Alam:2005pb,Alcaniz:2006ay}) and it will also be used in what
follows.

The measurement of the parameter $A$ (equation (\ref{ameas1}))
permits the addition of one more term in the $\chi^2$ of equations
(\ref{chi2}), (\ref{chi2cmb}) to be minimized with respect to $H(z)$
model parameters. This term is \be
\chi^2_{BAO}(\omm,w_0,w_1)=\frac{(A(\omm,w_0,w_1)-0.469)^2}{0.017^2}
\label{chi2bao} \ee

\subsection{The Cluster Baryon Gas Mass Fraction}

The baryon gas mass fraction for a range of redshifts can also be
used to constrain cosmological models $H(z)$. The basic assumption
underlying this method is that the baryon gas mass fraction in
clusters \cite{Arnaud:2005vq,Allen:2002sr,Allen:2004cd}: \be
f_{gas}=\frac{M_{b-gas}}{M_{tot}}\ee is constant, independent of
redshift and is related to the global fraction of the universe
$\frac{\Omega_b}{\omm}$ in a simple way. This relation may be
written as \be b
\frac{\omb}{\omm}=\frac{M_b}{M_{tot}}=(1+\alpha)\frac{M_{b-gas}}{M_{tot}}=(1+\alpha)f_{gas}
\label{omfgas}\ee where $b$ is a bias factor suggesting that the
baryon fraction in clusters is slightly lower than for the universe
as a whole. Also $1+\alpha$ is a factor taking into account the fact
that the total baryonic mass in clusters consists of both X-ray gas
and optically luminous baryonic mass (stars), the latter being
proportional to the former with proportionality constant
$\alpha\simeq 0.19\sqrt{h}$ \cite{Allen:2004cd}. Assuming now that
the hot gas in a cluster follows a spherically symmetric isothermal
model \ie \be n_e=n_{e0} (1+\frac{r^2}{r_c^2})^{-3\beta /2} \ee
where $\beta$ is a constant, $n_e$ is the electron number density
and $n_{e0}$, $r_c$ are the central electron density and the core
radius respectively, it is straightforward to show that
\cite{Sasaki:1996ss,Pen:1996sb} \be
M_{gas}(<R)=B(T_e,R/r_c,\beta)r_c^{3/2}L_X(<R)^{1/2}\label{mgas} \ee
In eq. (\ref{mgas}) $T_e$ is the electron temperature, $L_X(<R)$ is
the bolometric luminosity within radius $R$ and $B$ is a constant
independent of cosmological parameters. The quantities $r_c$,
$L_X(<R)$ are obtained from the observed angular core radius
$\theta_c$ and the observed apparent luminosity within $\theta$
($l_X(<\theta)$) as \ba
L_X(<R)&=& 4\pi d_L(z)^2 l_X(<\theta) \\
r_c&=&\theta_c d_A(z) \\
R&=&\theta d_A(z) \ea where $d_L (z)$ and $d_A(z)$ are the
luminosity and diameter distances which depend on the cosmological
model $H(z)$ as \be d_L(z)=(1+z)^2
d_A(z)=(1+z)\int_0^z\frac{dz'}{H(z')} \label{dlz2}\ee in the context
of a flat cosmology. From eqs (\ref{mgas})-(\ref{dlz2}) we obtain
\be M_{gas}(<R)=C(\theta_c,\theta,l_X,z)d_A(z)^{5/2}\label{mgasda}
\ee where all cosmological model dependence is encoded in $d_A(z)$
while C depends on observables characterizing each cluster
($\theta_c$, $l_X(<\theta)$,$T_e$,$z$).

\begin{table}[t]
\caption{\label{tablegas}Cluster data from Ref.
\cite{Allen:2004cd}.}
\begin{center}
\begin{tabular}{|c|c|c|}
  \hline
\;\;\;\;\;\;\; $ z_i $ \;\;\;\;\;\;\;&\;\;\;\;\;\;\; $ f_{gas,i} $ \;\;\;\;\;\;\;& \;\;\;\;\;\;\;$ \sigma_{gas,i}^2 $ \;\;\;\;\;\;\;\\
  \hline
 0.078  &  0.189  &  0.011  \\
 0.088  &  0.184  &  0.011  \\
 0.143  &  0.167  &  0.019  \\
 0.188  &  0.169  &  0.011  \\
 0.206  &  0.180  &  0.015  \\
 0.208  &  0.137  &  0.018  \\
 0.240  &  0.163  &  0.009  \\
 0.252  &  0.164  &  0.012  \\
 0.288  &  0.149  &  0.017  \\
 0.313  &  0.169  &  0.010  \\
 0.314  &  0.175  &  0.023  \\
 0.324  &  0.177  &  0.018  \\
 0.345  &  0.173  &  0.019  \\
 0.352  &  0.189  &  0.025  \\
 0.363  &  0.159  &  0.017  \\
 0.391  &  0.159  &  0.024  \\
 0.399  &  0.177  &  0.017  \\
 0.450  &  0.155  &  0.019  \\
 0.451  &  0.137  &  0.009  \\
 0.461  &  0.129  &  0.019  \\
 0.461  &  0.156  &  0.034  \\
 0.494  &  0.094  &  0.025  \\
 0.539  &  0.135  &  0.011  \\
 0.686  &  0.155  &  0.018  \\
 0.782  &  0.100  &  0.016  \\
 0.892  &  0.114  &  0.021  \\
  \hline
\end{tabular}
\end{center}
\end{table}

Similarly, the total cluster mass within $R$ $M_{tot}(<R)$ may be
obtained assuming that the intracluster mass is in hydrostatic
equilibrium as \be M_{tot}(<R)=D\; R=D\; \theta \; d_A(z)
\label{mtotr}\ee where $D$ is independent of cosmological model
parameters. From eqs (\ref{mgasda}) and (\ref{mtotr}) we find that
\be f_{gas}=\frac{M_{gas}(<R)}{M_{tot}(<R)}=\frac{C}{D \theta
}d_A(z)^{3/2}\equiv Q d_A(z)^{3/2} \label{fgasq}\ee where $Q$
depends only on individual cluster observables. Assume now that the
quantities $Q_i$ $(i=1,...,N)$ have been obtained for an observed
sample of $N$ clusters at redshifts $z_i$. Using eqs. (\ref{omfgas})
and (\ref{fgasq}) we have \be b\frac{\omb}{\omm}=(1+\alpha)
f_{gas}(z_i)=(1+\alpha)Q_i d_A(z_i)^{3/2}\label{fgaszi} \ee Define
now \be f_{gas}^{SCDM}(z_i)\equiv Q_i\;
d_A^{SCDM}(z_i)^{3/2}\label{fgascdm} \ee where $d_A^{SCDM}(z_i)$ is
the angular diameter distance corresponding to SCDM (flat $\omm=1$)
used as a reference model. Solving (\ref{fgascdm}) for $Q_i$ and
substituting in eq. (\ref{fgaszi}) we find \be
f_{gas}^{SCDM}(z_i)\equiv
\frac{b}{1+\alpha}\frac{\omb}{\omm}\left(\frac{d_A^{SCDM}(z_i)}{d_A(z_i)}\right)^{3/2}
\label{fgbas}\ee Since now $f_{gas}^{SCDM}$ is known observationally
(see eq. (\ref{fgascdm})) we can use eq. (\ref{fgbas}) to find the
best fit $d_A(z_i)$ and therefore the corresponding best fit
cosmological model.  We use the 26 cluster data for $f_{gas}^{SCDM}
(z_i)$ published in Ref. \cite{Allen:2004cd} and minimize
$\chi_{CBF}^2$ (Cluster Baryon Fraction) defined as \be
\chi^2_{CBF}(\omm,w_0,w_1)\equiv
\sum_{i=1}^{26}\frac{(f_{gas}^{SCDM}(z_i)-f_{gas,i})^2}{\sigma_{f_{gas,i}}^2}
\label{chi2cbf}\ee where $f_{gas}^{SCDM}(z_i)$ is given by eq.
(\ref{fgbas}) and the observed $f_{gas,i}$, $\sigma_{gas,i}^2$ used
in our analysis are shown in Table \ref{tablegas}. We treat
$\frac{b}{1+\alpha}\frac{\omb}{\omm}$ as a nuisance parameter and we
marginalize over it as follows: Define $\lambda \equiv
\frac{b}{1+\alpha}\frac{\omb}{\omm}$ then \be f_{gas}^{SCDM}(z_i)=
\lambda \cdot
\left(\frac{d_A^{SCDM}(z_i)}{d_A(z_i)}\right)^{3/2}\equiv \lambda
\cdot \tilde{f}_{gas}^{SCDM}(z_i)\label{lambda1}\ee and eq.
(\ref{chi2cbf}) after expanding with respect to $\lambda$ becomes
\be \chi^2_{CBF}\equiv \lambda^2 A-2\lambda B+ C\label{chi2l1}\ee
where \ba
A&=&\sum_{i=1}^{N} \frac{\tilde{f}_{gas}^{SCDM}(z_i)^2}{\sigma_{f_{gas,i}}^2} \label{bb} \nn \\
B&=&\sum_{i=1}^{N} \frac{\tilde{f}_{gas}^{SCDM}(z_i) \cdot f_{gas,i}}{\sigma_{f_{gas,i}}^2} \label{bb} \nn \\
C&=&\sum_{i=1}^{N}\frac{f_{gas,i}^2}{\sigma_{f_{gas,i}}^2 }
\label{cc} \ea Equation (\ref{chi2l1}) has a minimum for
$\lambda=B/A$ at \be \tilde{\chi}^2_{CBF}\equiv C-B^2/A
\label{chi2l2}\ee which is independent of $\lambda$. The
$\tilde{\chi}^2_{CBF}$ of eq. (\ref{chi2l2}) is the one actually
used in our analysis. It should be pointed out however that the
relative errors of these datapoints are more than $10\%$ and
therefore variations of $H(z)$ with given $\omm$ prior affect the
value of $\chi_{CBF}^2$ much less than SnIa, CMB and BAO data.

\subsection{Linear Growth Rate at z=0.15}

The 2dF galaxy redshift survey (2dFGRS) has measured the two point
correlation function at an effective redshift of $z_s=0.15$. This
correlation function is affected by systematic differences between
redshift space and real space measurements due to the peculiar
velocities of galaxies. Such distortions are expressed through the
{\it redshift distortion parameter} $\beta$ which connects the power
spectrum in redshift space $P_s ({\vec k})$ with the true galaxy
power spectrum $P_g(k)$ as \be P_s({\vec k})=(1+\beta \mu^2)^2
P_g(k)\label{psm} \ee where $\mu=\cos\theta$ and $\theta$ is the
angle between ${\vec k}$ and the line of sight. The parameter
$\beta$ can be observationally determined from the observed power
spectrum (or its Fourier transform, the correlation function) in
redshift space and may be shown to be connected to the growth factor
$D(a)$ at the effective redshift of the power spectrum
as\cite{Hamilton:1997zq} \be \beta = \frac{g}{b}\label{bet1} \ee
where \be g \equiv a \frac{dD(a)/da}{D(a)}\label{fdef1} \ee and $b$
is the bias factor ($b\equiv \frac{\delta_g}{\delta} $) which
connects the overdensity in galaxies $\delta_g$ with the matter
overdensity $\delta$. The derivation of eq. (\ref{bet1}) may be
sketched as follows\cite{Hamilton:1997zq}: The radial position of a
galaxy with low redshift $z$ and no peculiar velocity is
approximated by \be \chi(z)=\frac{z}{H_0}\ee and therefore its
redshift space position vector is \be {\vec
x}_s=\frac{z}{H_0}(\sin\theta \cos\phi,\sin\theta
\cos\phi,\cos\theta) \ee Due to gravitational effects however,
galaxies have peculiar velocities ${\vec v}$ which affect the
redshift $z$ as $z=H_0 x+{\vec v}\cdot {\hat x}$ where ${\hat x}$ is
the line of sight direction. Therefore, the true comoving distance
$x$ to the galaxy is connected to the redshift inferred distance as
\be \frac{z}{H_0}\equiv x_s=x+\frac{{\vec v}\cdot {\hat
x}}{H_0}\label{zx1} \ee Our goal is to connect the galaxy
overdensity in redshift space $\delta_s=\frac{\delta \rho_s ({\vec
x}_s)}{{\bar\rho}}$ with the corresponding galaxy overdensity
$\delta_g=\frac{\delta \rho_g (x)}{{\bar\rho}}$ in real space.

The number of galaxies in a spatial region is the same,
independent of the coordinate system \ie \be n_s({\vec x}_s)
d^3x_s=n({\vec x})d^3 x \ee or \be {\bar n}(1+\delta_s)={\bar n}
(1+\delta_g)\frac{dx}{dx_s}\frac{x^2}{x_s^2}\label{nd1} \ee where
${\bar n}$ is the average number density of galaxies. Assuming low
velocities and focusing on modes $kx>>1$ we obtain from eqs
(\ref{zx1}) and (\ref{nd1}) \be \delta_s({\vec x})=\delta_g({\vec
x})-\frac{\partial}{\partial x}(\frac{{\vec v}(x)\cdot {\hat
x}}{H_0}) \label{ds1} \ee By Fourier transforming eq. (\ref{ds1})
and expressing the peculiar velocity in terms of the mass
overdensity $\delta$ using the continuity equation \be
a\frac{d\delta}{dt}+i k v=0 \ee which
implies\cite{Hamilton:1997zq} \be {\vec v}(k,a)=\frac{i g a H
\delta(k,a){\vec k}}{k^2} \ee we find (for low redshifts $z<<1$)
\be {\tilde \delta}_s({\vec k})=\delta_g(k)+g\mu^2 \delta(k)
\label{delk}\ee where $\mu=\cos\theta ={\hat k}\cdot{\hat x}$ and
$g$ is defined by eq (\ref{fdef1}). We now use the bias factor
$b\equiv \frac{\delta_g}{\delta}$ to write eq. (\ref{delk}) as \be
{\tilde \delta}_s({\vec k})=(1+\frac{g}{b}\mu^2)\delta_g (k) \ee
which leads to eqs (\ref{psm}) and (\ref{bet1}).

The parameter $\beta=\frac{g}{b}$ may be measured from redshift
surveys by measuring $P_s({\vec k})$ and expanding both sides of
eq. (\ref{psm}) in Legendre polynomials ${\cal
P}_l(\mu)$\cite{Hamilton:1997zq}: \ba P_s({\vec k})&=&
P_s^{(0)}(k){\cal P}_0(\mu)+P_s^{(2)}(k){\cal
P}_2(\mu)+P_s^{(4)}(k){\cal P}_4(\mu)=\nn \\ &=&(1+\beta \mu^2)^2
P_g(k)=[(1+\frac{2}{3}\beta + \frac{1}{5}\beta^2){\cal
P}_0(\mu)+\nn \\ &+& (\frac{4}{3}\beta + \frac{4}{7}\beta^2){\cal
P}_2(\mu)+\frac{8}{35}\beta^2{\cal P}_4(\mu)]P_g(k) \ea Therefore
one way to measure $\beta$ is to use the measured quadrupole
$P_s^{(2)}(k)$ and monopole $P_s^{(0)}(k)$ as \be
\frac{P_s^{(2)}(k)}{P_s^{(0)}(k)}=\frac{\frac{4}{3}\beta +
\frac{4}{7}\beta^2}{1+\frac{2}{3}\beta + \frac{1}{5}\beta^2} \ee
which leads to the value of $\beta$ and (if the bias is
determined) to the value of $g$ at the effective redshift of the
redshift survey.

The correlation function (the Fourier transform of the galaxy
power spectrum) can also be used instead of the power spectrum to
obtain $\beta$ using a very similar method as the one described
above. Such a method was used along with others in Ref.
\cite{Verde:2001sf} to find \be \beta =0.49\pm 0.09 \ee at the
effective redshift of $z=0.15$ of the 2dF redshift survey. This
result may now be combined with the linear bias parameter \be
b=1.04\pm 0.11 \ee obtained from the skewness induced in the
bispectrum of the 2dFRGS\cite{Hawkins:2002sg} by linear biasing to
find the growth factor $g$ at $z=0.15$ as \be
g=a\frac{D'(a)}{D(a)}\vert_{z=0.15}=b\cdot \beta=0.51\pm 0.11 \ee
Such a result can in principle be used in combination with eq.
(\ref{greq}) with $f(k,a)=1$ to constrain $H(z)$ in the context of
general relativity or, if $H(z)$ is determined by other
observations, to determine the function $f(k,a)$ in eq.
(\ref{greq}) thus providing a robust test of alternative gravity
theories.

Under the assumption of general relativity, the measurement of the
Perturbations Growth Rate (PGR) $g(z=0.15)$ can be used to add one
more term to $\chi^2$ as \be
\chi^2_{PGR}(\omm,w_0,w_1)=\frac{(g(\omm,w_0,w_1)-0.51)^2}{0.11^2}
\label{chi2grr} \ee where $g(\omm,w_0,w_1)$ is obtained by solving
equation (\ref{greq}) with $f(k,a)=1$ and initial conditions
$D(a)\simeq a$ for $a\simeq 0$.
%\begin{figure}
%\hspace{0pt}\rotatebox{0}{\resizebox{0.5\textwidth}{!}{\includegraphics{fig7.eps}}}
%\vspace{-10pt}\caption{The best fit form of $w(z)$ for each
%dataset category for both $\Omega_{0m}=0.2$ and $\Omega_{0m}=0.3$
%along with the $1 \sigma$ errors (shaded region).} \label{fig7}
%\end{figure}

Notice however that the large relative error of this datapoint
(about $25\%$) makes the contribution of $\chi^2_{LSS}$ rather
insensitive to variations of $H(z)$. If however future observations
reduce the relative error of this measurement it is potentially very
important as it is a dynamical test capable of distinguishing
between general relativity and extended gravity theories along the
lines discussed in section I.

\subsection{Cosmological Data and PDL Crossing}

In order to identify the trends encoded in the current
cosmological data with respect to dark energy evolution and in
particular PDL crossing we group the data in three categories:
\begin{enumerate} \item SnIA Gold sample \item SnIA SNLS \item
Other Dark Energy Probes (ODEP) that include the CMB, BAO, Cluster
Baryon Fraction (CBF) and Perturbations Growth Rate (PGR) at
$z=0.15$
\end{enumerate} The corresponding $\chi^2$ to be minimized for
each category are

\begin{itemize}\item $\chi^2_{Gold}(\omm,w_0,w_1)$
(equation (\ref{chi2}))

\item $\chi^2_{SNLS}(\omm,w_0,w_1)$ (equation (\ref{chi2}))

\item Using equations (\ref{chi2cmb}), (\ref{chi2bao}),
(\ref{chi2cbf}) and (\ref{chi2grr}) we have \be \chi_{ODEP}^2 \equiv
\chi^2_{CMB} +\chi^2_{BAO}+\chi^2_{CBF}+ \chi^2_{PGR} \ee
\end{itemize}

In order to identify the dependence of the resulting best fits on
the $\omm$ prior used we have not marginalized over $\omm$. Instead
we have fixed $\omm$ and considered two cases ($\omm=0.2$ and
$\omm=0.3$). %%
The range between the two cases includes the current best fit value
of $\omm $ based on WMAP and SDSS which is\cite{Tegmark:2006az}
$\omm=0.24\pm 0.02$.%%
The errors of the best fit $w(z)$ were obtained using the covariance
matrix (see eg \cite{press92,Alam:2004ip}). The best fit form of
$w(z)$ for each dataset category is shown in Fig. 7 for both
$\omm=0.2$ and $\omm=0.3$. The corresponding $\chi^2$ contours in
the $w_0-w_1$ parameter space is shown in Fig. 8.

The following comments can be made based on our results summarized
in Figs. 7 and 8:
\begin{itemize}
\item The Gold dataset mildly favors dynamically evolving dark
energy crossing the PDL at $z\simeq 0.2$ over the cosmological
constant while the SNLS does not. \item Dark energy probes other
than SnIa mildly favor crossing of the PDL for low values of $\omm$
($\omm\lsim 0.25$) while for $\omm\simeq 0.3$ this trend is
significantly reduced. \item The best probes of dark energy with
currently existing data, other than SnIa are the CMB shift parameter
and the BAO peak. \end{itemize} The last point emerges by plotting
the third column of Figs. 7, 8 utilizing only CMB and BAO data. Such
plots are practically unchanged compared to those of Figs. 7, 8.

An important issue that needs to be addressed before closing this
section is the issue of systematics introduced by using a particular
parametrization to fit the cosmological expansion history. There is
no doubt that the best fit forms to a given dataset may differ
significantly between different parametrizations. However, the
particular property of PDL crossing at best fit has been shown to be
a robust feature for several $H(z)-w(z)$ parametrizations in the
context of the Gold dataset provided that these parametrizations
allow for PDL crossing. This was demonstrated in Ref
\cite{Lazkoz:2005sp} (Fig. 1) where we compared the best fit forms
of a wide range parametrizations to the Gold dataset. \clearpage
\newpage
\begin{center}
\begin{figure}[t!]
\vspace{1cm}{\resizebox{.9\textwidth}{!}{\vspace{1cm}
\hspace{5cm}\includegraphics{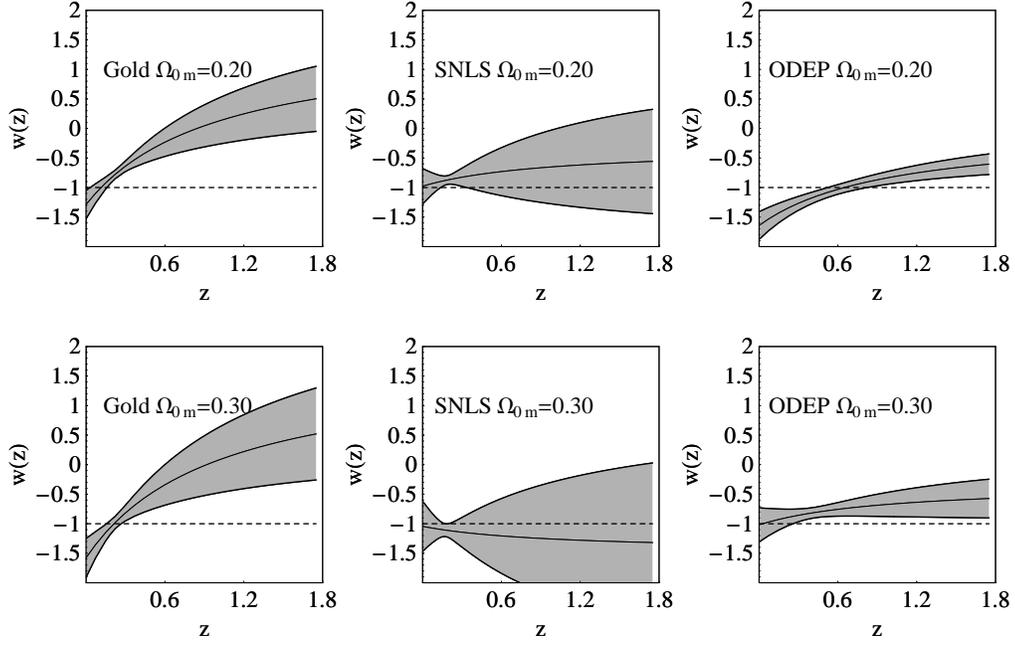}}}
\begin{minipage}{\textwidth}\caption{The best fit form of $w(z)$ for each
dataset category for both $\Omega_{0m}=0.2$ and $\Omega_{0m}=0.3$
along with the $1 \sigma$ errors (shaded region). The categories
are: Gold dataset (column 1), SNLS (column 2) and Other Dark Energy
Probes (ODEP column 3).}
\end{minipage}\label{fig7}
\end{figure}

\begin{figure}[b!]
\vspace{-1cm}{\resizebox{.9\textwidth}{!}{\vspace{-1cm}
\hspace{5cm}\includegraphics{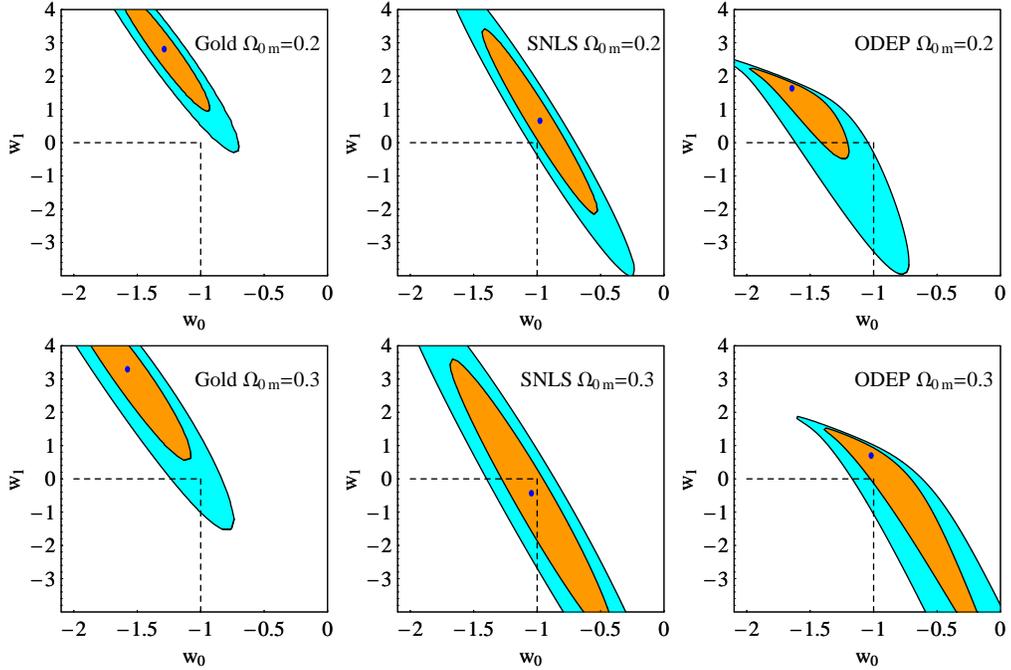}}}
\begin{minipage}{\textwidth}\caption{The $68\%$ and $95\%$ $\chi^2$ confidence
contours in the $w_0 - w_1$ parameter space for each dataset
category for both $\Omega_{0m}=0.2$ and $\Omega_{0m}=0.3$. Notice
that for the SNLS dataset $\Lambda$CDM is within the $1 \sigma$
region.} \end{minipage}\label{fig8}
\end{figure}

\end{center}
\clearpage In that paper it was demonstrated that even though the
high redshift properties at best fit were different for each
parametrization, the low redshift properties and in particular the
PDL crossing which appears at $z\simeq 0.2$ are very robust and
consistent among the different parametrizations. This robustness can
also be checked in the context of Other Dark Energy Probes.
%In Fig. 9 we show the best fit $w(z)$
%in the context of the following parameterizations \ba w(z) &=& ...
%\label{par1} \\ w(z) &=& ... \label{par2} \\ w(z) &=& ...
%\label{par3} \ea
We have fitted three different $H(z)$ parametrizations to the ODEP
along the lines of Ref. \cite{Lazkoz:2005sp}. We found that the best
fit forms for each parametrization indicate crossing of the PDL for
low $\omm$ but in this case there is a spread in the redshift range
of crossing in the range $0.4< z < 1.8$. In view of the very broad
redshift covering ($0<z<1089$) of the dominant probe of this class
which is the CMB shift parameter, the above spread in the crossing
range is not a significant indicator of systematics.

Another source of systematics that deserves attention is the
robustness of our results shown in Figs 7 and 8 with respect to the
theory of gravity considered. Modified gravity theories are expected
to affect dynamical tests of dark energy (which depend on the way
the metric couples to energy momentum tensor) but not geometrical
tests which detect directly the evolution of the FRW-metric. Tests
based on standard candles (SnIa) and standard rulers (CMB shift
parameter, BAO peak and X-ray gas mass fraction) are geometric tests
and we do not expect modified gravity theories to have a direct
effect on them. For example even though the CMB spectrum will vary
in modified gravity theories (in particular its ISW part), its
geometric features like the first peak location depend only on the
background metric and act as a standard ruler measuring directly the
integral of H(z).

On the other hand the test based on the growth rate of perturbations
at $z=0.15$ would give a different form for $H(z)$ in the context of
modified gravity theories (see Fig. 1). This effect however would
hardly modify our results of Figs. 7, 8 because the $1\sigma$ error
of the perturbations growth contributions to $\chi^2$ is much larger
that the geometrical tests contributions. In fact we have found that
even if we completely ignore the dynamical test contribution to
Figs. 7, 8 these figures remain practically unchanged. We should
stress however that there is an indirect effect of modified gravity
theories on SnIa standard candles. The possibility of a time varying
Newtons constant would induce a variation of the Chandrasekhar mass
and lead to a variation of the SnIa absolute luminosity. This effect
was studied in Ref. \cite{Nesseris:2006hp} for an evolving Newton's
constant consistent with nucleosynthesis and solar system bounds. It
was found that the effect is negligible for current SnIa data but
may have to be taken into account for future data with SNAP level
accuracy. %%

It is therefore clear that even though currently available dark
energy probes agree on the fact that the dark energy equation of
state $w(z)$ is close to $-1$ there is no universal trend with
respect to the evolution properties of dark energy and the existing
trends appear to depend on the value of $\omm$ prior considered.

\section{Conclusion-Outlook}

We have pointed out the importance of identifying observational
signatures that could distinguish between dark energy and extended
gravity theories as possible origins of the accelerating expansion
of the universe. The crossing of the PDL $w=-1$ can be identified as
such a signature based on a geometrical dark energy probe. We have
noted that the simplest theoretically motivated class of theories
that is consistent with such crossing is the class of extended
gravity theories. A representative example of this class of theories
includes scalar-tensor theories.

Even though there are currently no clear observational indications
for a PDL crossing, there are probes of accelerating expansion that
mildly favor such crossing over a constant equation of state. These
probes include the Gold SnIa dataset for {\it any} reasonable prior
for $\omm$ as well as the combination of other than SnIa probes
(CMB, BAO, Clusters Baryon Fraction and growth rate of
perturbations) with a low $\omm$ prior ($\omm\lsim 0.25$). On the
other hand a non-evolving $w\simeq -1$ appears to be favored by the
SNLS first year data.

Another (perhaps more robust) signature which could rule out dark
energy in favor of extended gravity theories is obtained by
utilizing the measurement of the growth rate of perturbations at
various redshifts. This test utilizes a `dynamical' (as opposed to
`geometrical') dark energy probe. We have demonstrated how can this
measurement be used to distinguish between the two classes of
theories. We have also pointed out that current 2dFGRS data are
consistent with $\Lambda$CDM in the context of general relativity
and require no extended gravity theory.

We have also discussed the ISW effect as another potentially
interesting dynamical probe of dark energy which is sensitive to the
growth rate of perturbations at recent redshifts. We have
demonstrated that in the context of general relativity and for fixed
shift parameter ${\bar R}$ and fixed $\omm$ the ISW effect on the
CMB power spectrum and the growth factor of perturbations $D(a)$
appear to be insensitive to modifications of the dark energy
equation of state. Degeneracies of the CMB power spectrum with fixed
shift parameter ${\bar R}$ have been discussed previously
\cite{Melchiorri:2002ux} but in those cases $\omm$ was varied
simultaneously with a $w(z)=w_0=const$ to keep ${\bar R}$ constant.
Thus, the large scale part of the CMB spectrum was seen to vary in
those studies due to the variation of $D(a)$ (ISW effect). The
degeneracy observed here can be used as an additional observational
discriminator between general relativity and extended gravity
theories. For example if the ISW effect is found to differ from the
expected form obtained by the measured shift parameter ${\bar R}$
and $\omm$ then this could be viewed as an indication of extensions
of general relativity. A more quantitative form of this argument is
an interesting potential extension of this work.%%
Modifications of the ISW effect are expected for example in theories
predicting a time varying Newton's constant.  In particular in Ref.
\cite{Sawicki:2005cc} it was shown that in the DGP model at late
times, perturbations enter a DGP regime in which the effective value
of Newton's constant increases as the background density diminishes.
This leads to a suppression of the ISW effect, bringing DGP gravity
into slightly better agreement with WMAP data than conventional
LCDM.%%

The  PDL crossing viewed as a geometrical signature of extended
gravity assumes the non-existence of phantom degrees of freedom
which is a reasonable assumption in view of the theoretical
problems of such degrees of freedom. On the other hand the
dynamical signature (growth rate of perturbations) assumes that
the dark energy perturbations can be ignored on sub-horizon scales
and can not mimic the effects of modified gravity theories
expressed through the function $f(k,a)$ (equation (\ref{greq})).
These assumptions make it important to identify further types of
observational signatures that can be used in combination with the
above, providing more robust tests that could distinguish between
the two classes of theories.

{\bf Numerical Analysis:} Our numerical analysis was performed using
a Mathematica code implementing $\chi^2$ fitting of multiple dark
energy probes and a modified version of CAMB allowing for a
dynamical $w(z)$ using the CPL parametrization. All the codes and
data along with detailed instructions are available at {\bf
http://leandros.physics.uoi.gr/pdl-cross/pdl-cross.htm }
\section*{Acknowledgements}
We thank L. Amendola, D. Polarski for useful discussions and M.
Reinecke for his help with the CMB numerical analysis. We also thank
S. Allen for providing the data of the cluster gas mass fraction
analysis (Table II) and R. Lazkoz for pointing out a couple of
important typos in the previous version of the paper. We acknowledge
the use of the Legacy Archive for Microwave Background Data Analysis
(LAMBDA). Support for LAMBDA is provided by the NASA Office of Space
Science. This work was supported by the program PYTHAGORAS-1 of the
Operational Program for Education and Initial Vocational Training of
the Hellenic Ministry of Education under the Community Support
Framework and the European Social Fund and by the European Research
and Training Network MRTPN-CT-2006 035863-1 (UniverseNet). SN
acknowledges support from the Greek State Scholarships Foundation
(I.K.Y.).

\end{document}